\documentclass[twocolumn,resetfootnote,longbib]{aastex7}
\usepackage{makecell}
\usepackage{booktabs}
\usepackage{amsmath}
\usepackage{amssymb}
\usepackage{ulem}
\usepackage{gensymb}

\newcommand\subs[1]{\textsubscript{#1}}
\newcommand\sups[1]{\textsuperscript{#1}}

\received{September 29, 2025}
\revised{March 20, 2026}
\accepted{April 1, 2026}
\submitjournal{PSJ}
\shorttitle{}
\shortauthors{Proudkii et al.}

\begin{document}


\title{Monitoring Volatile Evolution in Disrupting Comet D/2021 A1 (Leonard) with NOEMA and APEX}

\correspondingauthor{Timothy N. Proudkii}
\email{timproudkii@caltech.edu}

\author[0009-0005-1065-4109]{Timothy N. Proudkii}
\affiliation{Division of Physics, Mathematics, and Astronomy, California Institute of Technology, Pasadena, CA 91125, USA}
\affiliation{Solar System Exploration Division, NASA Goddard Space Flight Center, Greenbelt, MD 20771, USA}
\affiliation{Department of Physics, Virginia Tech, Blacksburg, VA 24061, USA}
\email{timproudkii@caltech.edu}

\author[0000-0002-6006-9574]{Nathan X. Roth}
\affiliation{Solar System Exploration Division, NASA Goddard Space Flight Center, Greenbelt, MD 20771, USA}
\affiliation{Department of Physics, American University, Washington, DC 20016, USA}
\email{nathaniel.x.roth@nasa.gov}


\author[0000-0002-1545-2136]{Jérémie Boissier}
\affiliation{Institut de Radioastronomie Millim\'{e}trique, Saint-Martin-d'H\`{e}res, F-38406, France}
\email{}

\author[0000-0002-8130-0974]{Dominique Bockelée-Morvan}
\affiliation{LIRA, Observatoire de Paris, Universit\'{e} PSL, CNRS, Sorbonne Universit\'{e}, Universit\'{e} Paris Cit\'{e}, Meudon, 92195, France}
\email{}

\author[0000-0003-2414-5370]{Nicolas Biver}
\affiliation{LIRA, Observatoire de Paris, Universit\'{e} PSL, CNRS, Sorbonne Universit\'{e}, Universit\'{e} Paris Cit\'{e}, Meudon, 92195, France}
\email{}

\author[0000-0001-6752-5109]{Steven B. Charnley}
\affiliation{Solar System Exploration Division, NASA Goddard Space Flight Center, Greenbelt, MD 20771, USA}
\email{}

\author[0000-0001-7694-4129]{Stefanie N. Milam}
\affiliation{Solar System Exploration Division, NASA Goddard Space Flight Center, Greenbelt, MD 20771, USA}
\email{}

\author[0000-0001-8233-2436]{Martin A. Cordiner}
\affiliation{Solar System Exploration Division, NASA Goddard Space Flight Center, Greenbelt, MD 20771, USA}
\affiliation{Department of Physics, The Catholic University of America, Washington, DC 20064, USA}
\email{}

\author[0000-0001-8843-7511]{Michael A. DiSanti}
\affiliation{Solar System Exploration Division, NASA Goddard Space Flight Center, Greenbelt, MD 20771, USA}
\email{}

\author[0000-0002-6391-4817]{Boncho P. Bonev}
\affiliation{Department of Physics, American University, Washington, DC 20016, USA}
\email{}

\author[0000-0002-8379-7304]{Neil Dello Russo}
\affiliation{Space Department, Johns Hopkins University Applied Physics Laboratory, Laurel, MD 20723, USA}
\email{}




\begin{abstract}
We report a pre-perihelion survey of volatile emissions from comet D/2021 A1 (Leonard) with the Northern Extended Millimeter Array (NOEMA; UT 2021 Nov. 5, 21, and Dec. 1) and the Atacama Pathfinder Experiment (APEX; UT 2021 Dec. 9-10), spanning heliocentric distances ($r_H$) from 1.3 to 0.80 au. We securely detected HCN and CS and place 3$\sigma$ upper limits on CH$_3$OH, H$_2$CO, and CO abundances. Line kinematics and NOEMA spatial constraints indicate that HCN was released at or near the nucleus (parent scale length $<300$ km), while CS showed higher gas expansion velocities and mixing ratios that increased with decreasing $r_H$---consistent with production from a distributed source. Across our campaign, CS mixing ratios relative to H$_2$O \added{increased} by a factor of $\sim$5, from \added{$0.02 \pm 0.01\%$  at $r_H$ = 1.3 au to $0.10\pm0.02\%$} by $r_H$ = 0.80 au. HCN mixing ratios in our data rose modestly, from \added{$0.04 \pm 0.02\%$} at $r_H$ = 1.3 au to \added{$0.07 \pm 0.02\%$} by $r_H$ = 0.81 au. \added{However, contemporaneous measurements from other facilities placed HCN consistently at a higher absolute level ($\sim\!0.08\%$) with additional variability. Once cross-facility measurements were included, the HCN abundance showed no statistically robust monotonic dependence on $r_H$.} Variability in both species during the mid-December outbursts and fragmentation suggests that D/2021 A1's volatile evolution reflected not only solar insolation but also disruption processes, underscoring the value of multi-epoch, multi-instrument monitoring to capture rapid, species-dependent changes.
\end{abstract}


\section{Introduction}\label{sec:intro}
\subsection{Comets as Primitive Solar System Bodies}
Primarily composed of dust, rock, and various ices, comets are some of the most primitive bodies in the solar system. Comets formed during the era of planet formation and their initial composition should provide a glimpse into the thermal and chemical processes present in the cold disk midplane where they formed \citep{Bockelee2004}. After their formation, the migration of the giant planets scattered comets into their present-day major dynamical reservoirs: the Oort cloud or the Kuiper disk \citep{Gomes2005,Morbidelli2005,Levison2011}. Here, in the cold, desolate outskirts of the solar system, comets remained relatively unaltered due to minimal thermal processing or gravitational self-heating. Within this context, comets are often thought of as ``fossils'' of the early solar system. As comets chart their way through the inner solar system (heliocentric distances (\textit{r}\subs{H}) $<$ 3 au), they release volatiles and form their coma due to increased solar insolation. Among various methods of investigation, remote sensing of the coma provides one of the primary means to infer the composition of the comet's nucleus---offering invaluable insights into nascent solar system conditions. 

Cometary observations are commonly analyzed using the Haser model \citep{Haser1957}, which categorizes gases as ``parent'' (directly subliming from nucleus ices, indicating its composition) or ``product'' (formed through photolysis or other processes in the coma). A wide range of observations, spanning from microwave to UV wavelengths, have provided valuable insights into the physical and chemical properties of the regions in the protoplanetary disk where comets formed. For example, the relatively high abundance of the hypervolatile carbon monoxide (CO) in comets \citep{Mumma2011a, DelloRusso2016} indicates they likely formed in the outer solar system owing to its low sublimation temperature. On the other hand, the detection of crystalline silicates in some comets implies that materials processed at high temperatures and small heliocentric distances were incorporated into their nuclei \citep{Bregman1987,Zolensky2006}. Coupled with recent dynamical modeling (e.g., \citealt{Levison2011,Nesvorny2017}), this juxtaposition of high- and low-temperature materials in comets suggests the presence of large-scale mixing in the protoplanetary disk and a ``mixed'' comet formation region. Additionally, volatile abundances in solar system comets provide insights into the ice-phase midplane chemistry in protoplanetary disks around other stars, serving as one of the few observational constraints on ice-phase disk models \citep{Willacy_2022}. As such, measurements of cometary volatiles provide critical tests of models describing the physical and chemical conditions in disks---helping us to refine our understanding of solar system formation.

\subsection{Motivation and Study of Comet D/2021 A1 (Leonard)}
Previous observations of comets at small \textit{r}\subs{H} have shown intriguing results, especially during critical stages near perihelion. Among these comets, D/2012 S1 (ISON) emerged as a point of interest due to its close encounter with the Sun and eventual disruption. Cometary studies with state-of-the-art interferometers---such as the Atacama Large Millimeter Array (ALMA) and the Northern Extended Millimeter Array (NOEMA)---provide spatial-spectral maps of coma volatiles at high sensitivity and avenues to explore dynamic coma processes. For example, \cite{Cordiner2017} revealed significant variability in the production of gas and dust as ISON approached the Sun using ALMA. These temporally and spatially resolved observations suggested that hydrogen isocyanide (HNC) likely originated from thermal degradation of organic refractory material (i.e., a ``product'' volatile)---shedding light on a topic that has been the subject of active debate within the field of cometary science \citep{Irvine1998,Rodgers2005,Lis2008,Cordiner2014}.

Studies of ISON at other wavelengths revealed that mixing ratios of volatiles such as hydrogen cyanide (HCN) and formaldehyde (H$_2$CO) increased with decreasing heliocentric distance at small \textit{r}\subs{H} \citep{Irvine2004,DiSanti2016}. Interestingly, this behavior is not unique to ISON. For instance, in a study of 30 comets between 1997 and 2013, \cite{DelloRusso2016} found that the mixing ratios of ammonia (NH$_3$), H$_2$CO, and acetylene (C$_2$H$_2$) may also increase at small heliocentric distances once a thermal threshold is reached: \textit{r}\subs{H} $<$ 0.8 au. This trend suggests that there are common underlying mechanisms governing the production and release of these volatiles as comets approach the Sun. The increase in mixing ratios at small \textit{r}\subs{H} likely results from the enhanced sublimation of volatile ices and the subsequent release of gas and dust particles from the comet's nucleus. Furthermore, there is the possibility of additional sources influencing the observed trends, particularly related to the thermal decomposition of organic refractory materials \citep{Irvine2004,fray2006heliocentric,DelloRusso2016,DiSanti2016,Cordiner2017}. Carbon monosulfide (CS) provides a clear example: its heliocentric dependence cannot be explained alone by carbon disulfide (CS$_2$) photolysis, implying a distributed refractory source \citep{Cottin2008}. This interpretation is consistent with Rosetta's detection of organo-sulfur molecules alongside ammonium salts at comet 67P/Churyumov-Gerasimenko, which may serve as stable reservoirs that release sulfur-bearing species into the coma \citep{Altwegg2020}.

This study examines the early stages of these heliocentric trends by cataloging the inbound behavior of volatile release in comet D/2021 A1 (Leonard). The comet's small perihelion distance ($\sim\!0.6$ au), high predicted peak brightness \citep[m$_1\sim 4$;][]{Biver2024}, and close approach to Earth made it subject of a coordinated global observing campaign and an ideal candidate for such an investigation. Documenting the evolution of inbound coma activity allows us to inform our view of the fundamental processes governing volatile release and coma formation before the onset of intense thermal-driven activity near perihelion. 

Comet D/2021 A1 (Leonard; hereafter A1), discovered by Greg J. Leonard on UT 2021 January 3, was a returning long period comet with a semi-major axis a $\sim$2,040 au (corresponding to P $\sim $80,000 yr; Nakano Note\footnote{\url{https://www.oaa.gr.jp/~oaacs/nk/nk4621.htm}}). A1 made its nearest approach to Earth on UT 2021 December 12 at a distance of 0.233 au. On UT 2021 December 13, \cite{Crovisier2021} recorded a sudden increase of line intensities while monitoring A1's outgassing rate through the OH 18-cm transition with the Nan\c{c}ay radio telescope.

This abrupt enhancement marked the beginning of a series of recurrent outbursts that significantly boosted the comet's activity. Within a single day, the observed OH line intensities nearly doubled, with modeled OH production rates rising from $2.6 \times 10^{28}$ $\mathrm{molecules~s^{-1}}$ on UT 2021 December 12 to $6.3 \times 10^{28}$ $\mathrm{molecules~s^{-1}}$ on UT 2021 December 13. Activity continued to intensify, peaking at $22 \times 10^{28}$ $\mathrm{molecules~s^{-1}}$ by UT 2021 December 15. Production rates then declined to $9 \times 10^{28}$ $\mathrm{molecules~s^{-1}}$ by UT 2021 December 17, only to surge again, reaching $27 \times 10^{28}$ $\mathrm{molecules~s^{-1}}$ by UT 2021 December 19 \citep{Crovisier2021}.

Amidst these outbursts, A1 made a historically close approach to Venus on UT 2021 December 18, passing at just 0.028 au---the closest known cometary approach to Venus \citep{Zhang_2021}. During this period of heightened activity, sometime in mid-December 2021, A1 began to disintegrate, with a pre-disintegration nucleus radius of 0.6 $\pm$ 0.2 km \citep{Jewitt_2023}. \cite{Jewitt_2023} suggested that this disintegration was most likely caused by rotational instability driven by outgassing torques. 

Exactly one year after its discovery, A1 reached perihelion on UT 2022 January 3, with a perihelion distance of q = 0.6151 au. Following perihelion, \cite{2022ATel15189....1J} reported another significant outburst between UT 2022 January 6 and 8, during which A1's production rates increased by a factor of four. These observations provide valuable insights into the volatile dynamics and structural integrity of dynamically active long-period comets nearing the end of their life cycle. For these reasons, A1 was an exceptional target for studying the interplay between volatile production, nucleus fragmentation, and outburst mechanisms in dynamically evolving comets.

Here we present an investigation of coma volatiles in the disrupting comet A1. \added{In this context, we use the term ``disrupting comet'' to describe a nucleus undergoing sustained structural instability, characterized by recurrent outbursts, rapidly evolving production rates, fragmentation, and eventual disintegration. In contrast to ``typical'' active comets, whose volatile release generally follows relatively smooth heliocentric trends driven by solar insolation, disrupting comets may expose fresh volatile-rich layers and release refractory material more readily. As a result, their observed mixing ratios can depart from steady heliocentric compositional trends. Our pre-perihelion monitoring observations with NOEMA and the Atacama Pathfinder EXperiment (APEX), obtained prior to A1's fragmentation, therefore provide a temporal record of volatile activity immediately preceding the onset of large-scale structural instability.} 

\added{We securely detected molecular emission from HCN and CS in the A1's coma. Additionally, we derived stringent (3$\sigma$) upper limits for methanol (CH\subs{3}OH), H\subs{2}CO, and CO mixing ratios relative to H$_2$O. Our analysis focuses on the evolution of volatile mixing ratios with decreasing heliocentric distance and on constraining the production mechanisms of HCN and CS. The majority of our observations were conducted when A1 ranged from $r_H$ = 1.3 au to 0.80 au, at which point H$_2$O was subliming vigorously and dominating coma activity. At these distances, previous studies suggest that mixing ratios should remain relatively constant \citep{DelloRusso2016}. However, because A1 approaches the $\sim$0.8 au thermal threshold during our final epochs, our datasets probe the onset of any emerging compositional deviations within this regime.}

In Section~\ref{sec:obs}, we present a detailed account of our observations, including the instrumental setups. In Section~\ref{sec:data_reduction}, we outline the data reduction process. Section~\ref{sec:modeling_methods} describes the modeling techniques used. Section~\ref{sec:results} presents our results and analysis. In Section~\ref{sec:discussion}, we interpret our findings in the context of other comet A1 observations.

\section{Observations} \label{sec:obs}
\subsection{NOEMA Observations}\label{subsec:NOEMA_obs}
We conducted pre-perihelion observations toward A1 on UT 2021 November 5, November 21, and December 1, using the NOEMA interferometer in the 10C array configuration. We utilized the PolyFiX wideband correlator with the Band 3 receiver (frequency coverage 196-276 GHz). Observations were carried out in two spectral configurations, each tuned to a different central frequency to cover multiple molecular transitions simultaneously. The first spectral configuration, observed on 2021 November 5 and 21, covered a frequency range ($\sim$245-269 GHz) that sampled HCN ($J$=3-2) and multiple CH\subs{3}OH transitions ($J_K$ = $J_3$ - $J_2$). The second spectral configuration, observed on November 5, November 21, and December 1, covered a range ($\sim$224-247 GHz) that included CH\subs{3}OH ($J_K$=$5_K-4_K$), CS ($J$=5-4), H\subs{2}CO ($J_{Ka,Kc}$=3$_{1,2}$-2$_{1,1}$), and CO ($J$=2-1). We report both the interferometric data and ON-OFF position switching data obtained during these observations. For tracking the position of the comet, we used JPL HORIZONS ephemerides (JPL \#29). Weather conditions during our observations varied, with an average PWV of 0.5 mm - 0.7 mm, 3 mm, and 4 mm on November 5, November 21, and December 1, respectively. The sources 3C84, 1156+295, 0851+202, LKHA101, 2010+723, and J1257+324 were used as calibrators. Spectra were converted to the main-beam brightness temperature scale using the forward and beam efficiencies provided by NOEMA for the dates of observation. We obtained a native frequency resolution of 62.4 kHz.

The acquisitions were performed by repeating a sequence of acquisitions of about 30 minutes which included: pointing and focusing, performing cross correlation scans on the calibration sources, taking ON-OFF measurements on the comet (2 minutes) and then performing cross correlation scans on the target source (30 scans of 45 seconds for a total of 22.5 minutes). The ON-OFF measurements involve observing the target for one minute and then switching to a reference position located 300 arcseconds in azimuth from the target for one minute. These measurements allowed us to measure the total power of the electromagnetic radiation from the target and subtract any background or instrumental \added{signal}. These ON-OFF observations are essentially autocorrelations taken by the NOEMA antennas. The cross-correlation scans measure the correlation between the signals received by different pairs of antennas and can be used to reconstruct an image of the target and extract spatial information (discussed in section~\ref{visibility_modeling_methods}). The observing log for NOEMA ON-OFF observations and for interferometric observations are shown in Table~\ref{tab:NOEMA_all}.

\begin{deluxetable*}{cccccccccc}
\tablecaption{NOEMA Observations (ON-OFF above; Interferometric below)}\label{tab:NOEMA_all}
\tablewidth{0pt}
\tablehead{
\colhead{UT Date} & \colhead{UT Time} & \colhead{\textit{t}\subs{int}} &
\colhead{\textit{r}\subs{H}} & \colhead{$\Delta$} & \colhead{$\phi$} &
\colhead{\textit{N}\subs{ants}} & \colhead{Baselines} & \colhead{$\theta$\subs{INT}} & \colhead{PWV} \\
\colhead{} & \nocolhead{} & \colhead{(minutes)} & \colhead{(au)} &
\colhead{(au)} & \colhead{($\degr$)} & \colhead{} & \colhead{(m)} &
\colhead{($\arcsec$)} & \colhead{(mm)}
}
\startdata 
\multicolumn{10}{c}{\textbf{[ON--OFF]}} \\[2pt]
\multicolumn{10}{c}{$\nu$ = 235.5 GHz, $\theta$\subs{AC} = 21.4$\arcsec$} \\
\hline
2021 Nov. 5  & 6:30-10:20 & 14.00 & 1.3  & 1.4  & 42   & 10 & \dots & \dots & 0.7 \\
2021 Nov. 21 & 6:30-11:45 & 19.80 & 1.08 & 0.88 & 59.4 & 9  & \dots & \dots & 3   \\
2021 Dec. 01 & 3:00-7:38  & 18.00 & 0.93 & 0.52 & 80.1 & 8  & \dots & \dots & 4   \\
\hline
\multicolumn{10}{c}{$\nu$ = 257.1 GHz, $\theta$\subs{AC} = 19.6$\arcsec$} \\
\hline
2021 Nov. 5  & 3:57-5:10  & 6.00  & 1.3  & 1.4  & 42   & 10 & \dots & \dots & 0.5 \\
2021 Nov. 21 & 2:18-4:00  & 7.80  & 1.08 & 0.88 & 59.4 & 9  & \dots & \dots & 3   \\
\hline
\multicolumn{10}{c}{\textbf{[Interferometric]}} \\[2pt]
\multicolumn{10}{c}{$\nu$ = 235.5 GHz, $\theta$\subs{AC} = 21.4$\arcsec$} \\
\hline
2021 Nov. 5  & 6:30-10:45 & 55.20 & 1.3  & 1.4  & 42   & 10 & 23-204 &  1.1$\times$0.6          & 0.7 \\
2021 Nov. 21 & 6:30-12:15 & 93.60 & 1.08 & 0.88 & 59.4 & 9  & 28-207 &  1.1$\times$0.6          & 3   \\
2021 Dec. 1  & 3:00-8:00  & 79.20 & 0.93 & 0.88 & 59.4 & 8  & 28-207   &  1.5$\times$0.8          & 4   \\
\hline
\multicolumn{10}{c}{$\nu$ = 257.1 GHz, $\theta$\subs{AC} = 19.6$\arcsec$} \\
\hline
2021 Nov. 5  & 4:00-5:33  & 20.40 & 1.3  & 1.4  & 42   & 10 & 23-204 & 1.0$\times$0.6 & 0.5 \\
2021 Nov. 21 & 2:21-4:24  & 43.80 & 1.08 & 0.88 & 59.4 & 9  & 28-207 & 1.0$\times$0.6 & 3
\enddata
\tablecomments{\textit{t}\subs{int} is the total on-source integration time. 
\textit{r}\subs{H}, $\Delta$, and $\phi$ are the heliocentric distance, geocentric distance, and phase angle (Sun-comet-Earth) at the time of observations. 
\textit{N}\subs{ants} is the number of antennas used; for interferometric rows we give the baseline range. 
$\theta$\subs{INT} is the synthesized-beam FWHM for interferometric observations.
PWV is the mean precipitable water vapor at zenith. 
In the sub-headers, $\theta$\subs{AC} is the autocorrelation beam FWHM at $\nu$ (the mean frequency of the instrumental setting).
Cells marked \dots\ are not applicable.}
\end{deluxetable*}

\subsection{Synergy Between NOEMA ON-OFF and Interferometric Data}\label{subsec:NOEMA_obs_syn}
Our NOEMA interferometric data are sensitive to spatial structures ranging from $\sim\!$ 1$\arcsec$ to $\sim\!$ 10$\arcsec$, beyond which emission becomes resolved out due to the lack of short baselines in the interferometric array. During the observations, the baselines ranged from approximately 20 m to 200 m, providing sensitivity to spatial scales of $\sim\!$ 1$\arcsec$ at the observing frequency (see Table~\ref{tab:NOEMA_all}). For a distance 1 au away from Earth, 1$\arcsec$ subtends a projected distance of 725 km. Structures smaller than $\sim\!$ 1$\arcsec$ are unresolved because they fall below the spatial resolution limit set by the observing frequency and the maximum baseline length. 

On the other hand, our NOEMA ON-OFF data contain flux from the full NOEMA primary beam, which spans $\sim\!21.4\arcsec$ at 235.5 GHz. Given that cometary comae can span hundreds of arc-seconds \citep{cordiner2023gas}, the ON-OFF data contain more flux per beam when compared to the interferometric data. For these reasons, the ON-OFF observations are particularly useful for measuring the total flux and large-scale structure of the cometary coma. Additionally, since these ON-OFF spectra are of higher S/N, they are particularly valuable for precise characterization of spectral line profiles, which in turn ensures more robust constraints on the modeled gas kinematics.

The interferometric data, however, provide complementary strengths. Their higher spatial resolution allows for detailed investigation of localized features within the coma, such as jets or anisotropies in outgassing. Additionally, interferometric visibilities can be modeled to constrain spatial parameters such as the molecular parent scale length, providing insights into the spatial extent of molecular emission and distinguishing between nucleus-originated and coma-originated species. In this framework, ``parent'' species are released directly from the nucleus or near-surface sublimation, ``daughter'' species are generated in the coma by photolysis of gas-phase parents, and ``distributed'' source species are those whose extended spatial distributions cannot be explained by gas-phase processes alone and likely require release from refractory material.

\subsection{APEX Observations}\label{subsec:APEX_obs}
In addition to these NOEMA observations, we conducted pre-perihelion observations towards A1 on UT 2021 December 9 and 10 using the Atacama Pathfinder Experiment (APEX) 12-meter single-dish radio telescope using the Fast Fourier Transform Spectrometer (FFTS). On December 9, we observed using the SEPIA345 receiver (frequency coverage 272-376 GHz) which allowed us to target molecular emission from HCN ($J$=4-3), CS ($J$=7-6), H\subs{2}CO ($J_{Ka,Kc}$=5$_{1,5}$-4$_{1,4}$), and CH\subs{3}OH ($J_K=7_K-6_K$). On December 10, we observed in using the nFLASH230 receiver (frequency coverage 196-281 GHz) which allowed us to target molecular emission from CS ($J$=5-4), CO ($J$=2-1), H\subs{2}CO ($J_{Ka,Kc}$=3$_{1,2}$-2$_{1,1}$), and CH\subs{3}OH ($J_K=5_K-4_K$). We display the observing log in Table~\ref{tab:APEX_obslog}. As with our NOEMA observations, we tracked A1 using JPL HORIZONS ephemerides (JPL \#29). Weather conditions during our observations varied. On December 9 we recorded an average PWV of 2 mm and on December 10 an average of 5 mm. The sources IRC+10216 and RX-Boo were used as calibrators. We converted the spectra to velocity space in the cometocentric rest frame. Additionally, we placed the fluxes onto the main beam scale using efficiencies cataloged on the APEX website\footnote{https://www.apex-telescope.org/telescope/efficiency/} using main-beam efficiencies of 0.7 for the December 9 epoch and 0.77 for December 10. For these observations, we obtained a native frequency resolution of 61 kHz.

\begin{deluxetable*}{cccccccccc}
\tablecaption{APEX Observing Log}\label{tab:APEX_obslog}
\tablewidth{0pt}
\tablehead{
\colhead{Setting} & \colhead{UT Date} & \colhead{UT Time} & \colhead{\textit{t}\subs{int}} &
\colhead{\textit{r}\subs{H}} & \colhead{$\Delta$} & \colhead{$\phi$} & \colhead{$\nu$}  & \colhead{$\theta_{mb}$} & \colhead{PWV} \\
\colhead{} & \colhead{} & \nocolhead{} & \colhead{(minutes)} & \colhead{(au)} &
\colhead{(au)} & \colhead{($\degr$)} & \colhead{(GHz)} & \colhead{($\arcsec$)} & \colhead{(mm)}
}
\startdata
1 & 2021 Dec. 09 & 11:46-14:02 & 21.60 & 0.81 & 0.27 & 124 & 353.3 & 18 & 2 \\
2 & 2021 Dec. 10 & 11:17-14:12 & 32.20 & 0.80 & 0.25 & 133 & 229.5 & 27 & 5
\enddata
\tablecomments{\textit{t}\subs{int} is the total on-source integration time. \textit{r}\subs{H}, $\Delta$, and $\phi$ are the heliocentric distance, geocentric distance, and phase angle (Sun-comet-Earth), respectively, of A1 at the time of observations. $\nu$ is the mean frequency of each instrumental setting, $\theta_{mb}$ is the approximate FWHM of the telescope's main beam, and PWV is the mean precipitable water vapor at zenith during the observations.}
\end{deluxetable*}

\section{Data Reduction}\label{sec:data_reduction}
From our NOEMA observations, we obtained both ON-OFF position-switching and interferometric data. Spectra from the ON-OFF observations were extracted using standard routines in \texttt{GILDAS/CLASS} (\citealt{Pety2005, team2013gildas}; \url{https://www.iram.fr/IRAMFR/GILDAS/}). To enhance the signal-to-noise ratio (S/N), the native spectral resolution was sparingly smoothed until an optimal balance between S/N and spectral detail was achieved. These processed spectra formed the basis for subsequent radiative transfer modeling. APEX observations followed an identical reduction workflow.

\begin{figure*}
    \centering
    \includegraphics[width=1\textwidth]{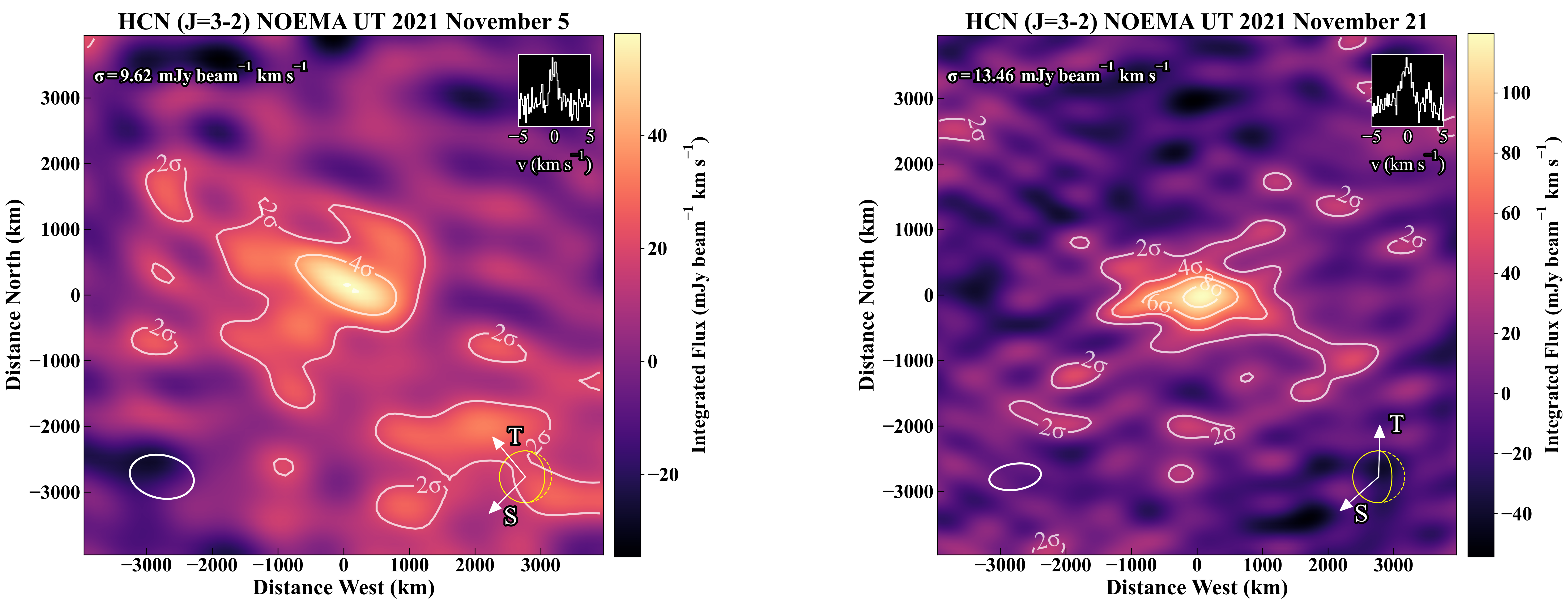}
    \caption{Spectrally integrated flux maps for NOEMA HCN ($J$=3-2) detections on November 5 (left) and November 21 (right). In the bottom left of the map, we display the size and orientation of the synthesized beam. The lower right of the map indicates the comet's illumination, the Sun's orientation (S), and the dust trail (T), while the upper right panel displays the spectral line extracted from the brightest pixel. For November 5, the rms noise is 9.62 mJy~beam$^{-1}$~km~s$^{-1}$, with the contours drawn from 2$\sigma$ to 4$\sigma$. For November 21, the rms noise is 13.46 mJy~beam$^{-1}$~km~s$^{-1}$, with the contours drawn from 2$\sigma$ to 8$\sigma$.}
    \label{fig: maps}
\end{figure*}

For the NOEMA interferometric data, visibility calibration and imaging were performed using the \texttt{CLIC} and \texttt{MAPPING} packages of the \texttt{GILDAS} software suite. Standard calibration procedures corrected for bandpass, phase, and flux variations using observations of standard calibrator sources, as described in Section~\ref{subsec:NOEMA_obs}. Imaging was carried out with natural visibility weighting, and the flux threshold was set to match the rms (root mean square) noise for each image. Prior to deconvolution, we interactively defined polygonal CLEAN masks to enclose regions of statistically significant emission. Deconvolution of the synthesized beam was then achieved using the Högbom algorithm. The resulting spectrally integrated flux maps of HCN ($J$=3-2) from November 5 and November 21 are shown in Figure~\ref{fig: maps}. 

\section{Radiative Transfer Modeling}\label{sec:modeling_methods}
To interpret the cometary spectra detected in our NOEMA and APEX observations, we used \texttt{SUBLIME}, a comprehensive and versatile tool designed to simulate the excitation and emission of molecular lines in cometary atmospheres \citep{Cordiner_2022, cordiner2023gas}. \texttt{SUBLIME} is a three-dimensional radiative transfer method capable of modeling both spectral line profiles and interferometric visibility data. \texttt{SUBLIME} includes pumping by solar radiation, collisions with H\subs{2}O and electrons, and non-LTE treatment of coma gases. Building on the principles of the \cite{Haser1957} model, \texttt{SUBLIME} incorporates a radial density profile, treating coma molecules as parent species, photochemical daughter species, or a combination of both, expanding outward at a constant velocity. For a more comprehensive overview of \texttt{SUBLIME}, refer to \cite{Cordiner_2022} and \cite{cordiner2023gas}. 

In our \texttt{SUBLIME} modeling, photodissociation rates for HCN, CO, H\subs{2}CO, and CH\subs{3}OH were adopted from \cite{Huebner2015}. CS photodissociation rates were adopted from \cite{biver2011molecular}. We used HCN-H\subs{2}O collisional rates from \cite{Dubernet2019} and CO-H\subs{2}O collisional rates from \cite{Faure2010}. We assumed that CS-H\subs{2}O, H\subs{2}CO-H\subs{2}O, and CH\subs{3}OH-H\subs{2}O collisional rates scaled as CS-H\subs{2} \citep{Lique2006}, H\subs{2}CO-H\subs{2} \citep{Wisenfeld&Faure2013}, and CH\subs{3}OH-H\subs{2} \citep{Rabli2010} respectively \citep[taken from the LAMDA database][]{Schoier2005}. We utilized water production rates measured in A1 from SOHO/SWAN data \citep{Combi2023}. \added{SOHO/SWAN measurements coincided closely with several observing epochs, with temporal offsets ranging from less than 24 hours to $\sim\!4$ days. We use the SWAN measurement from the nearest available date to each NOEMA/APEX epoch, without modification to the reported best-fit water production value. The SWAN measurements adopted here also occur outside the epochs or reported major outburst activity (around mid-December, starting on UT 2021 December 13; \citealt{Crovisier2021}). The SWAN time series nevertheless exhibits short-timescale variability (see Figure 3 of \citealt{Combi2023}). In addition, SWAN's large effective field of view samples emission over an extended region of the coma, which introduces temporal smoothing due to coma expansion across the aperture relative to the smaller APEX/NOEMA beams. As a result, the SWAN-derived water production rates may not exactly reflect the activity sampled by mm observations. To account for the combined effects of short-timescale variability and the small temporal offsets between SWAN and mm measurements, we include an additional 10$\%$ fractional uncertainty in the adopted SWAN H$_2$O production rates (added in quadrature). This term represents a first-order allowance for temporal mismatch between SWAN and our APEX/NOEMA measurements, reflecting day-to-day variability and imperfect simultaneity, as well as possible differences arising from the large-aperture nature of the SWAN observations.} 

\subsection{Model Geometry}\label{model_geometry}
To interpret the coma's spatial and spectral characteristics, we adopted a two-region asymmetric outgassing geometry within \texttt{SUBLIME}, following the methods of \cite{Cordiner_2022}. In this framework, the coma is divided into two distinct outgassing regions, $\Omega_1$ and $\Omega_2$, each with independent kinematical properties. This geometry allows us to capture solar-driven asymmetries in the coma while maintaining a physically motivated, minimally complex structure.

Each solid angle region ($\Omega_1, \Omega_2$) was assigned distinct parameters, including independent water production rates (Q$_1$, Q$_2$), outflow velocities ($v_{exp1}, v_{exp2}$), and radial temperature profiles ($T_1(r), T_2(r)$). Radial temperature profiles can be derived if multiple molecular transitions for a given species (such as CH$_3$OH) spanning a range of excitation energies are detected. However, we did not detect CH\subs{3}OH during our observations and, therefore, we were not able to constrain $T_1(r)$ and $T_2(r)$. Instead, we employed an isothermal temperature profile. In our November modeling, we used a kinetic temperature of 25 K, while in our December modeling, we used a temperature of 65 K. These kinetic temperature values were based on IRAM 30-m observations of A1 \citep{Biver2024}. 

In cases where the data support more localized or directional outgassing, we further parametrized the $\Omega_1$ region as a conical jet with a half-opening angle $\theta_{jet}$, oriented at a phase angle ($\phi$) relative to the observer and position angle ($\psi$) in the sky plane. The second region, $\Omega_2$, represents the broader remaining coma. When sufficient signal-to-noise was not available to constrain $\theta_{jet}$, we fixed the two regions to symmetric sunward and anti-sunward hemispheres ($\theta_{jet}$ = 90$\degree$).

This model framework provides a consistent and physically interpretable structure capable of reproducing both symmetric and asymmetric spectral line profiles. Applying the same formalism across all observation dates allows for direct comparison between retrieved gas properties and ensures that any temporal variations reflect real changes in the coma and not differences in modeling assumptions. 

\subsection{Visibility Modeling}\label{visibility_modeling_methods}
Using \texttt{SUBLIME}, we modeled the NOEMA interferometric visibilities to reconstruct the (projected) two-dimensional spatial distribution of molecular emission in the coma. Following \cite{cordiner2023gas}, the visibility modeling process uses Fourier-domain fitting to avoid imaging artifacts. In our approach, we utilized the \texttt{vis\_sample} program \citep{Loomis2018} to transform our \texttt{SUBLIME} model images into the Fourier domain using the same $uv$-coverage as our NOEMA observations. By fitting model visibilities, we can in principle constrain both symmetric and asymmetric features. In practice, we fit the full complex visibilities without $uv$-binning; directional information is retained but the limited S/N restricts robust sensitivity to asymmetries, so our modeling primarily constrains the radially averaged coma structure.

Our NOEMA ON-OFF and interferometric observations were taken contemporaneously in NOEMA's compact configuration. While the ON-OFF mode provides total flux measurements across the full primary beam, the compact interferometric data retain partial sensitivity to larger spatial scales---though not as fully as the ON-OFF observations---while offering improved resolution of finer coma structures near the nucleus. To leverage the strengths of both datasets, we simultaneously modeled the ON-OFF (incorporated as zero-spacing visibilities) and interferometric data using the two-region asymmetric jet outgassing model described in Section~\ref{model_geometry}. The choice of this model reflects the physical expectation of solar-driven asymmetries and localized jets in cometary comae while maintaining a geometry simple enough to be constrained by the data. Our combined ON-OFF + visibility modeling approach facilitates consistent retrieval of volatile mixing ratios relative to H$_2$O, expansion velocities, and geometric parameters across multiple spatial scales. Here, all retrieved parameters were constrained from fitting the interferometric visibilities, with the ON-OFF spectrum providing the zero-point spacing flux constraint. Similar modeling approaches have proven successful in \citealt{Roth2021a, cordiner2023gas}. 

To distinguish between parent, daughter, and distributed source species and to characterize the spatial distribution of molecular emission, we treat the molecular parent scale length ($L_p$) as a critical model parameter. $L_p$ describes the radial extent of parent molecules before processing into daughter/distributed species and provides insights into where molecules were produced in the coma. In our modeling framework, an $L_p$ of 0 km corresponds to a direct nucleus sublimation (parent model), with no secondary production in the coma. A non-zero but physically plausible $L_p$ represents a daughter model, in which a volatile progenitor photodissociates into a daughter species at a rate inversely proportional to $L_p$. In contrast, cases where the inferred $L_p$ cannot be explained by gas-phase photochemistry are treated as evidence for a distributed source---implying release from refractory material within the coma rather than from volatile sublimation alone.

Accurately constraining parent scale lengths requires a nuanced approach and consideration of multiple factors. First, molecular densities of cometary species exhibit an inverse exponential behavior with respect to radius. As a consequence, uncertainties associated with the parent scale lengths can display significant asymmetry. Second, our NOEMA observations were carried out in its compact configuration, which is ideal for capturing extended coma emission (e.g., H$_2$CO), but less capable of resolving molecules released very close to the nucleus. Therefore, to determine model parameters, we generated synthetic visibilities for a grid of parameter values, computed the $\chi^2$ statistic from the model-data comparison of the complex visibilities (with the ON-OFF spectrum providing the $uv = 0$ constraint), and performed a $\Delta\chi^2$ comparison relative to the minimum value. We divided the process into two steps: an initial exploration of broader parameters to identify approximate best-fit conditions and a refined analysis focused on $L_p$. Both steps use the combined interferometric + ON-OFF data for their analysis. 

\begin{figure*}
    \centering
    \includegraphics[width=1.0\textwidth]{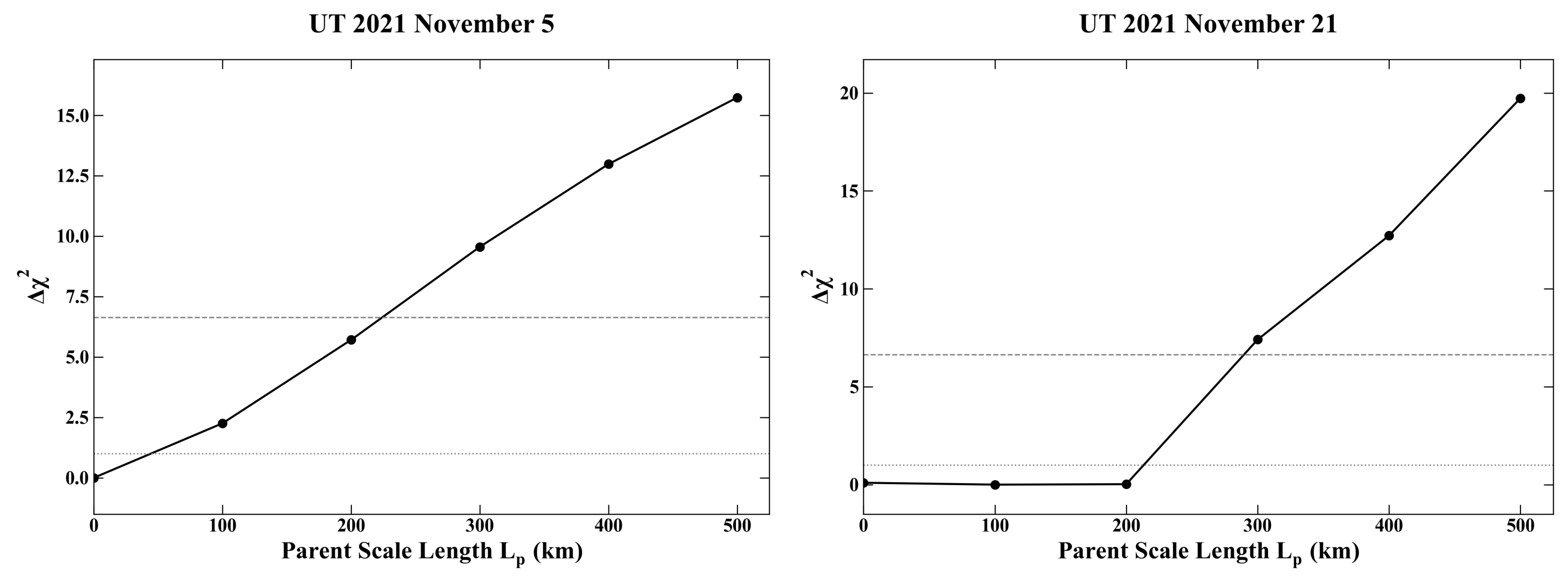}
    \caption{The two panels show $\Delta\chi^{2}$ as a function of parent scale length ($L_p$) for November 5 (left) and 21 (right). \textbf{The black line connects the discrete $\Delta\chi^{2}$ points to guide the eye.} Dotted horizontal lines indicate the $\Delta \chi^2 = 1$ threshold (1$\sigma$, 68$\%$ confidence), while dashed horizontal lines mark the $\Delta \chi^2 = 6.63$ threshold (2.6$\sigma$, 99$\%$ confidence), with $\chi^2$ computed from the difference between the observed interferometric and synthetic complex visibilities (real and imaginary parts) and the ON-OFF spectrum providing the $uv = 0$ constraint.}
    \label{fig: chi-analysis}
\end{figure*}

Initially, we constrained $Q_1/Q_2$ (the ratio of molecular production rates in the two regions), molecular abundance, $v_{exp1}$ and $v_{exp2}$ (hemisphere specific expansion velocities), and the half-open angle ($\theta_{jet}$). We performed model optimization for a range of $\theta_{jet}$, assuming that the jet was oriented along the Sun-comet line. For each value of $\theta_{jet}$, the $Q_1/Q_2,~v_{exp1},~v_{exp2},$ and molecular abundance parameters were allowed to freely vary. We then computed $\Delta\chi^2(\theta_{jet}) = \chi^2(\theta_{jet})-\chi^2_{min}$, with $\chi^2$ computed for each model iteration. The best-fit $\theta_{jet}$ corresponds to the minimum $\chi^2$, and confidence intervals were derived from the $\Delta\chi^2 = 1$ (68\%) and $6.63$ (99\%) thresholds. After this step, we fixed $\theta_{jet}$ and $Q_1/Q_2$ to their best-fit values and repeated the analysis but now for $L_p$, allowing $v_{exp1}$, $v_{exp2}$, and the daughter abundance to vary. Figure~\ref{fig: chi-analysis} presents the $\Delta\chi^2$ results for the $L_p$ analysis for HCN; \added{we discuss this figure in more detail in Section~\ref{subsec:NOEMA_results}}.

\subsection{Spectral Line Modeling}\label{spectral_modeling_methods}
For molecular species only detected with our ON-OFF spectra or with APEX, we used \texttt{SUBLIME}'s spectral line modeling capabilities to constrain gas kinematics and molecular abundances in the coma. Since the ON-OFF spectra are extracted from a single beam centered on the nucleus, they do not provide adequate spatially resolved information and we did not fit for $L_p$ in this framework. Instead, we adopted L$_p$ or parent photodissociation rates from values reported in the literature (e.g., ``effective'' photodissociation rates for H$_2$CO from \citealt{Roth2021a}), and focused our modeling on volatile mixing ratios relative to H$_2$O and expansions velocities. This approach allowed us to capture the bulk chemical composition and outflow properties without over-interpreting spatial structure that cannot be constrained by the available data alone. 

For our targeted species, CS and H$_2$CO both exhibited spatially extended production in cometary comae, and we modeled them accordingly. For H$_2$CO, although near-infrared measurements of the inner coma (e.g. \citealt{DelloRusso2016}) have sometimes been interpreted as evidence for direct nucleus release, in situ data from Giotto's Neutral Mass Spectrometer first established H$_2$CO as a distributed source in comet 1P/Halley \citep{Meier1993}. Additionally, subsequent millimeter-wave studies have tied H$_2$CO production to an unknown parent source \citep{Biver1999, Bockelee2000, Milam2006}. Recent high-resolution ALMA observations of the inner coma suggested scale lengths ranging from 1,000 km to 7,000 km, depending on heliocentric distance and gas expansion velocities \citep{Cordiner2014, Cordiner2017, Roth2021a, cordiner2023gas}. We adopted an ``effective'' parent photodissociation rate for H$_2$CO of $\beta = 1.17\times10^{-4}\mathrm{~s^{-1}}$ at $r_H=1$ au from \citet{Roth2021a}, scaling it with $r_H$. This photodissociation rate, together with the modeled expansion velocity, determined the effective parent scale length used in our analysis of H$_2$CO ($L_p\approx$ 7,000 km at 1 au). For CS, numerous comet studies have shown it to arise as a daughter species, with CS$_2$ suggested as a potential precursor, although some comets have shown a CS distribution which cannot be explained by gas-phase chemistry and instead require a refractory progenitor \citep{Feldman2004, Boissier2007, Bogelund2017, Roth2021a, biver2022observations}. Recent analyses have further demonstrated that the effective parent scale length for CS is typically 3-5 times larger than the CS$_2$ photodissociation scale length \citep{Roth2021a, biver2022observations, Biver2024}. For these reasons, we modeled CS as a distributed source and adopt an $L_p$ of 1,000 km for CS---identical to those used by \citet{Biver2024} in their contemporaneous analysis of comet A1.

On the other hand, we modeled HCN, CH$_3$OH, and CO as parent species. For HCN, millimeter-wave and interferometric observations consistently show compact emission coinciding with the inner coma \citep{Cordiner2014, Roth2021a, cordiner2023gas}. \added{Additionally, our own NOEMA visibility modeling (Figure~\ref{fig: chi-analysis}) places stringent upper limits on the HCN parent scale length, favoring $L_p \approx 0$ km and ruling out extended production on scales $\gtrsim$ a few hundred kilometers at 99$\%$ confidence. This compact spatial distribution is therefore consistent with direct nucleus sublimation}. Although CH$_3$OH is also released from the nucleus, its coma abundance could be enhanced by the sublimation of dirty ice grains and larger ice chunks lofted into the coma as seen in more ``hyperactive'' comets \citep{Bonev2021, Roth2021b, cordiner2023gas}. Nevertheless, interferometric maps reveal a predominately near-nucleus distribution and therefore we treated CH$_3$OH as a parent molecule. For CO, we treated it as a parent molecule because interferometric observations have shown CO emission to be consistent with direct nucleus outgassing \cite{Bockel_e_Morvan_2010}. However, we note that the origin of CO in comets is debated, with possible contributions from both direct nucleus release and sublimation of icy grains or extended sources \citep{eberhardt1988co, disanti1999identification, disanti2001carbon, brooke2003spectroscopy, disanti2003evidence}. In our case, since CO is only used to establish an upper limit, we adopted the parent designation as a simplifying assumption. 

\section{Results}\label{sec:results}
\subsection{NOEMA Results}\label{subsec:NOEMA_results}
We detected HCN ($J$=3-2) in both our NOEMA ON-OFF and interferometric observations. As described in Section~\ref{visibility_modeling_methods}, we modeled the combined visibility dataset, which includes both the NOEMA interferometric data and the ON-OFF data incorporated as short-baseline visibilities, as a function of $uv$-radius to constrain the HCN parent scale length. As part of this modeling, we tested a range of fixed half-open angle values and determined the best-fit half-open angle for each observation by minimizing the $\chi^2$ value. The best fit half-open angle was $60 \pm 5\degree$ for the November 5th observation and $55 \pm 3\degree$ for the November 21st observation. Our modeling efforts yielded 2.6$\sigma$ (99$\%$ confidence) upper limits on the HCN parent scale length. \added{As shown in Figure~\ref{fig: chi-analysis}, the November 5th NOEMA interferometric dataset places a strong constraint of $L_p <225$ km. In this case, the $\Delta \chi^2$ curve is minimized at the boundary $L_p = 0$~km and increases monotonically with $L_p$, indicating that the data strongly disfavor an extended source and that HCN is consistent with nuclear release. For the November 21st observation, the $\Delta \chi^2$ evaluations from 0-200 km show multiple comparable minima at low $L_p$ and only begin to rise at larger $L_p$. The 99$\%$ confidence threshold is crossed at $L_p \approx 295$~km, which we therefore adopt as an upper limit for HCN's $L_p$ on November 21st.}

\begin{figure*}
    \centering
    \includegraphics[width=1.0\textwidth]{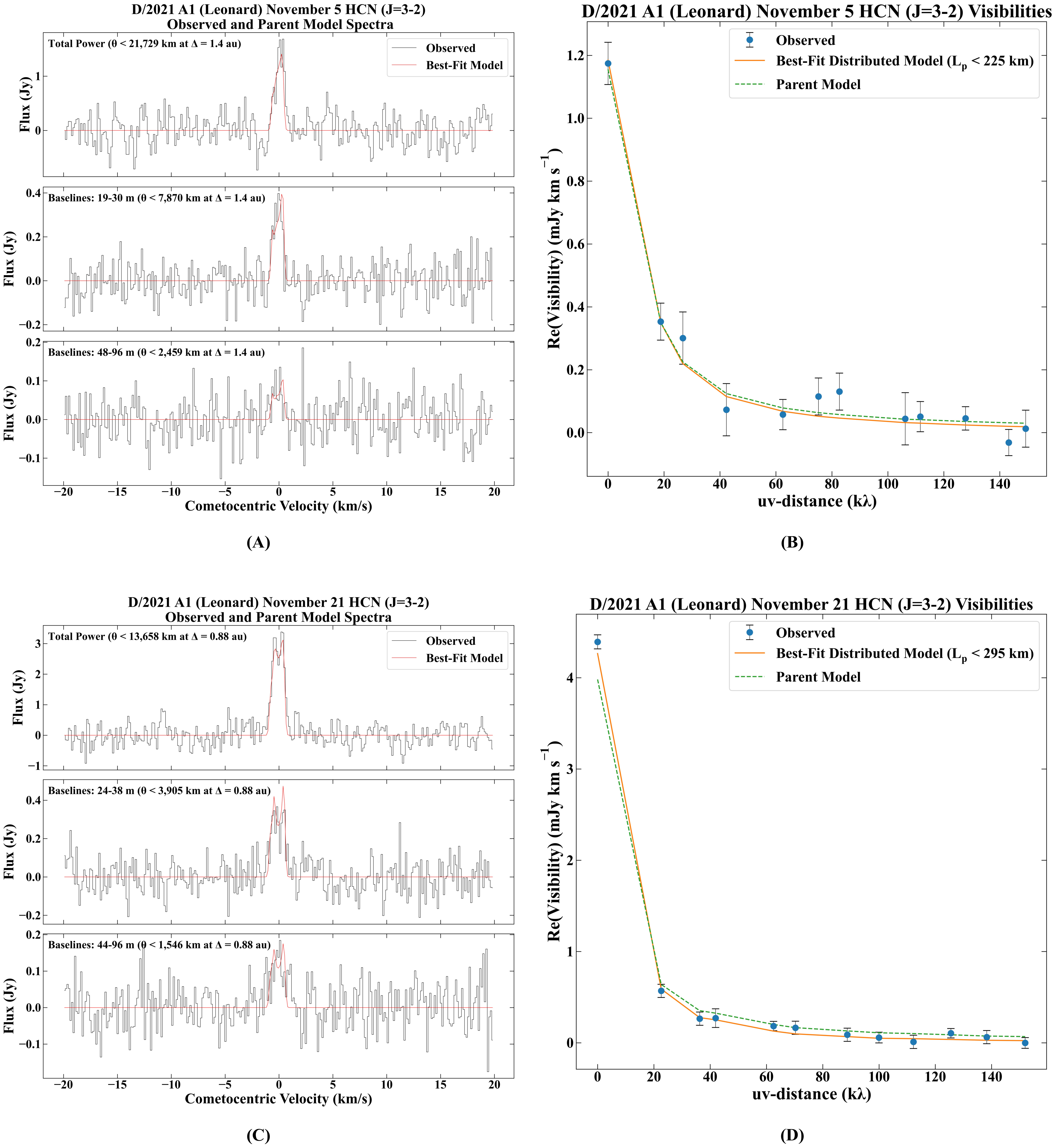}
    \caption{(A) HCN ($J$=3-2) spectrum of A1 on UT 2021 November 5 (black). The top panel shows the ON-OFF spectrum, while the lower two panels display spectra extracted over different baseline ranges (i.e., angular scales). The spectra are shown at a frequency resolution of 125 kHz (velocity resolution 0.15 km s$^{-1}$). The best-fit model is overplotted in red. (B) Real part of the observed visibility amplitude as a function of projected baseline length for HCN on UT 2021 November 5. The parent model and the upper-limit distributed source model are overplotted for comparison. The \textit{uv}-distance is given in units of k$\lambda$. (C-D) HCN spectra for November 21, with traces and labels as in (A-B).}
    \label{fig: HCN-comp}
\end{figure*}

In Figure~\ref{fig: HCN-comp}, we present both the spectral line profiles and the interferometric visibilities compared to our best-fit models. Panels A and C show the ON-OFF spectra alongside spectra extracted from different baseline ranges. The best-fit parent model is overplotted for comparison. Panels B and D display the real part of the visibility amplitudes as a function of $uv$-distance for November 5 and November 21, respectively. In both epochs, the observed visibilities closely follow the parent model; the inclusion of a distributed component (labeled ``Distributed Model'') provides no significant improvement to the fit. The November 5 dataset, in particular, shows excellent agreement between the parent-only and (upper limit) distributed models and places a stringent constraint on any possible extended source. The November 21 dataset allows for a somewhat larger parent scale length, but combined with the shallow $\Delta\chi^2$ minimum (Figure~\ref{fig: chi-analysis}) and consistency with the parent model, any distributed HCN production must be minor. These results show that the HCN coma of A1 is dominated by direct nucleus release.

In addition to HCN ($J$=3-2), we also detected CS ($J$=5-4) in the NOEMA ON-OFF observations (see Figure~\ref{fig:CS_plots}). We did not detect CS with our interferometric observations. Given the asymmetry in most CS line profiles, we modeled CS transitions with the two-region framework described in Section~\ref{model_geometry}, separating the sunward and anti‐sunward hemispheres equally with $\theta_{jet} = 90 \degree$. We did not fit for the half-opening angle here because the data did not provide sufficient signal-to-noise to constrain it reliably. Fixing it at 90$\degree$ avoids introducing a poorly constrained free parameter while remaining physically reasonable for solar illumination conditions and consistent with the observed line profiles.

\subsection{APEX Results}\label{subsec:APEX_results}
We securely detected molecular emission from HCN ($J$=4-3) and CS ($J$=7-6 and $J$=5-4) with APEX. The last panel of Figure~\ref{fig:HCN_plots} shows a detection of HCN on 2021 December 9 while the last row of Figure~\ref{fig:CS_plots} shows our CS detections with APEX. Line profiles of both CS and HCN were asymmetric with dominant blue components. Therefore, we employed the same two-region model used in our NOEMA spectral analysis. For our CS and HCN detections on December 9, the data were not able to constrain $\theta_{jet}$, so we modeled those transitions using the fixed-hemisphere case ($\theta_{jet} = 90\degree$). For our CS detection on December 10, given its sufficient signal-to-noise and strong asymmetry in the line shape, we allowed the jet half-opening angle ($\theta_{jet}$) to vary as a free parameter.

\subsection{Upper Limits on Mixing Ratios}

For non-detected transitions, we determined 3$\sigma$ upper limits on their coma mixing ratios relative to H$_2$O and on their production rates. For each line, the per-channel rms was propagated across the expected velocity width ($\sim$2 km/s) to calculate a 3$\sigma$ limit on the integrated line intensity. \added{For species with multiple observed transitions (e.g, H$_2$CO and CH$_3$OH), the treatment depended on the line distribution. For H$_2$CO, the transitions are widely separated in frequency and excitation; therefore, we modeled only the transition expected to be strongest under typical coma excitation conditions of A1 \citep{Biver2024}. In practice, we modeled the strongest ortho-H$_2$CO transition available and report the total H$_2$CO abundance assuming the standard ortho-to-para ratio of 3:1 (corresponding to the nuclear-spin statistical weight ratio). In contrast to H$_2$CO, CH$_3$OH has multiple comparably strong lines within the same rotational ladder and close in frequency. For CH$_3$OH we modeled the $J_K=7_K-6_K$ A species transitions in the 338.4-338.8~GHz range and report the total methanol abundance assuming an A:E ratio of 1:1. This assumption is commonly adopted in cometary millimeter analyses (e.g., \citealt{Bockelee1994, de2018measuring, bergman2022emission}) and follows the standard treatment that the A and E symmetry species are equally abundant under statistical equilibrium. For CO, only a single transition ($J$=2-1) was observed, and no additional symmetry or ladder treatment was required.} To calculate upper limits on mixing ratios relative to H$_2$O, we ran iterative forward models with SUBLIME, fixing the gas outflow geometry to the best-fit parameters determined from a detected transition on that same date, using HCN when available (owing to its higher signal-to-noise ratio) and CS otherwise. The model mixing ratio was varied until the predicted integrated intensity matched the observed 3$\sigma$ limit. Our APEX observations on December 9 and 10 provide the most stringent constraints, and we therefore restrict our reported upper limits to these dates. These results are summarized in Table~\ref{tab:apex_upperlimits}, which lists the molecule, transition, rest frequency, observation date, rms noise, 3$\sigma$ line area, and the derived 3$\sigma$ mixing ratio and production rate upper limits \added{(computed using the upper bound of our adopted $Q(\mathrm{H}_2\mathrm{O})$ value from \citealt{Combi2023})}.

\begin{deluxetable*}{lcccccccc}
\tablecaption{APEX 3$\sigma$ upper limits for non-detected species.\label{tab:apex_upperlimits}}
\tablehead{
\colhead{Molecule} & \colhead{Transition} & \colhead{Rest Frequency} & \colhead{Obs. Date} & \colhead{rms} & \colhead{3$\sigma$ Line Area} & \colhead{3$\sigma$ Upper Limit} & \colhead{3$\sigma$ Upper Limit}\\
\colhead{} & \colhead{} & \colhead{(GHz)} & \colhead{(2021)} & \colhead{(mK)} & \colhead{(mK km s$^{-1}$)} & \colhead{(mixing ratio)} & \colhead{(mol s$^{-1}$)}}
\startdata
CH$_3$OH & multiple  & 338.4 - 338.8 & Dec. 9  & 43 & 180 & $<1.2$\% & $<3.9\times10^{26}$\\
H$_2$CO  & $5_{1,5}$-4$_{1,4}$   & 351.769 & Dec. 9  & 44 & 61 & $<0.2$\% & $<6.4\times10^{25}$ \\
CO       & 2-1                   & 230.538 & Dec. 10  & 22 & 36 & $<7$\% & $<2.2\times10^{27}$ \\
\enddata
\tablenotetext{}{Note --- The CH$_3$OH upper limit is derived from a stack of transitions listed in Table~\ref{tab:ch3oh_lines}. Transition quantum numbers are labeled as $J'_{K'_a , K'_c} - J''_{K''_a, K''_c}$ for H$_2$CO and as $J'-J''$ for CO.}
\end{deluxetable*}

\begin{deluxetable}{lcc}
\tablecaption{CH$_3$OH transitions used in the stacked upper-limit analysis (A--CH$_3$OH, $v_t=0$). \label{tab:ch3oh_lines}}
\tablehead{
\colhead{Transition} & \colhead{Rest Frequency} & \colhead{$E_u$} \\
\colhead{} & \colhead{(GHz)} & \colhead{(K)}}
\startdata
$7_{0}-6_{0}\,A^{+}$        & 338.408698 & 65.0 \\
$7_{6}-6_{6}\,A^{-}$        & 338.442367 & 258.7 \\
$7_{-6}-6_{-6}\,A^{+}$      & 338.442367 & 258.7 \\
$7_{5}-6_{5}\,A^{-}$        & 338.486322 & 202.9 \\
$7_{-5}-6_{-5}\,A^{+}$      & 338.486322 & 202.9 \\
$7_{-4}-6_{-4}\,A^{+}$      & 338.512632 & 145.3 \\
$7_{4}-6_{4}\,A^{-}$        & 338.512644 & 145.3 \\
$7_{-2}-6_{-2}\,A^{+}$      & 338.512853 & 102.7 \\
$7_{3}-6_{3}\,A^{-}$        & 338.540826 & 114.8 \\
$7_{-3}-6_{-3}\,A^{+}$      & 338.543152 & 114.8 \\
$7_{2}-6_{2}\,A^{-}$        & 338.639802 & 102.7 \\
\enddata
\tablenotetext{}{Note --- Transition quantum numbers are labeled as $J'_{K'} - J''_{K''}$.}
\end{deluxetable}

\subsection{Summary of Results}\label{subsec:results}
\begin{figure*}
    \centering
    \includegraphics[width=1\textwidth]{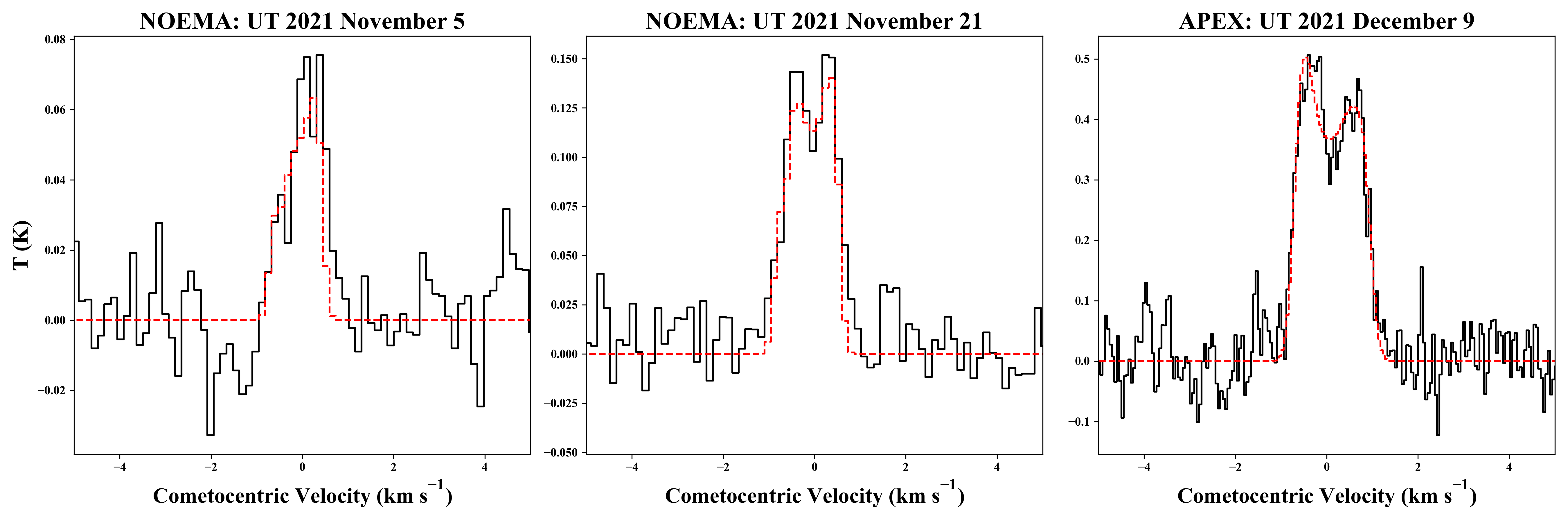}
    \caption{Evolution of the HCN spectral line in comet A1. The black line depicts the extracted spectra, while the red line represents our modeled data. NOEMA spectra shown here are from the ON-OFF observations.}
    \label{fig:HCN_plots}
\end{figure*}

\begin{figure*}
    \centering
    \includegraphics[width=1\textwidth]{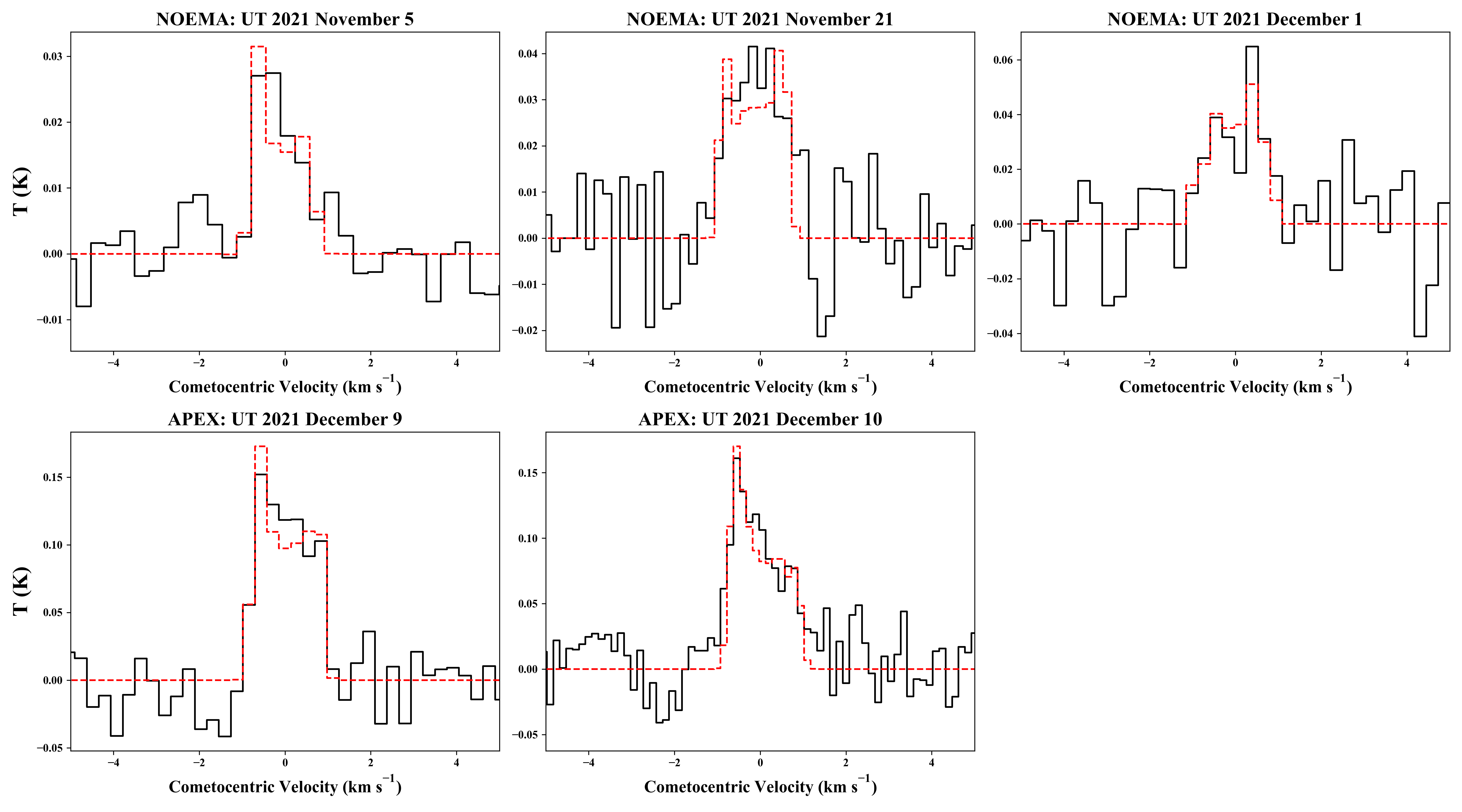}
    \caption{Evolution of the CS spectral line in comet A1. The black line depicts the extracted spectra, while the red line represents our modeled data. NOEMA spectra shown here are from the ON-OFF observations.}
    \label{fig:CS_plots}
\end{figure*}

Based on our APEX and NOEMA ON-OFF observations, we have successfully detected molecular emissions of HCN and CS. The evolution of the HCN line profile, along with our modeled line profile, is depicted in Figure~\ref{fig:HCN_plots}, while Figure~\ref{fig:CS_plots} illustrates the results for CS. The modeling results---including expansion velocities, jet half-open angles, and mixing ratios relative to water---are presented in Table~\ref{tab:composition_table}. In our modeling, we used H$_2$O production rates from \citet{Combi2023} \added{with an added $10\%$ fractional uncertainty (discussed in Section~\ref{sec:modeling_methods})}. To account for comparing measurements from different observatories, we included an additional 10$\%$ uncertainty associated with the absolute flux scale, which we applied to only flux-dependent quantities such as the derived mixing ratios. $Q(X)$ can be calculated based on our reported abundances and assumed $Q(\mathrm{H_2O})$ (listed in Table~\ref{tab:composition_table}). This makes it possible to directly compare our results with studies that report absolute $Q(X)$ values based on different assumptions for $Q(\mathrm{H_2O})$.

The most stringent 3$\sigma$ upper limits for CO, H\subs{2}CO, and CH\subs{3}OH across all epochs were obtained with APEX, as summarized in Table~\ref{tab:apex_upperlimits}. Our NOEMA interferometric observations place stringent (2.6$\sigma$) upper limits on HCN parent scale lengths, with an upper limit of $<225$ km from our November 5th observation (best-fit half-open angle: $60 \pm 5\degree$), and $<295$ km from our November 21st observation (best-fit half-open angle: $55 \pm 3\degree$).

\begin{deluxetable*}{ccccccccc}
\tablecaption{Molecular Composition and Geometry of D/2021 A1 (Leonard)}\label{tab:composition_table}
\tablewidth{0pt}
\tablehead{
\colhead{Date} & 
\colhead{$Q({\mathrm{H_2O}})$\,\sups{a}} &  
\colhead{$M_1$\,\sups{b}} &  
\colhead{$\int T_b~dv$\,\sups{c}} &  
\colhead{$v_{\mathrm{exp1}}$\,\sups{d}} & 
\colhead{$v_{\mathrm{exp2}}$\,\sups{e}} & 
\colhead{$Q(X) / Q({\mathrm{H_2O}})$\,\sups{f}} & 
\colhead{$Q_1 / Q_2$\,\sups{g}} &
\colhead{$\theta_{jet}$\,\sups{h}}\\
\colhead{(2021)} & 
\colhead{($10^{28}$ mol s$^{-1}$)} & 
\colhead{(km s$^{-1}$)} & 
\colhead{(K km s$^{-1}$)} & 
\colhead{(km s$^{-1}$)} & 
\colhead{(km s$^{-1}$)} & 
\colhead{(\%)} &
\colhead{} &
\colhead{($\degree$)}
}
\startdata
\multicolumn{9}{c}{CS} \\
\hline
Nov. 5$^\dagger$  & 2.34 $\pm$ 1.0 & -0.01 $\pm$ 0.14 & 0.036 $\pm$ 0.006 & 0.84 $\pm$ 0.03 & 0.69 $\pm$ 0.06 & 0.02 $\pm$ 0.01 & 2.9 $\pm$ 1.7 & 90\\
Nov. 21$^\dagger$ & 3.61 $\pm$ 0.5 & -0.09 $\pm$ 0.16 & 0.053 $\pm$ 0.008 & 0.93 $\pm$ 0.04 & 0.61 $\pm$ 0.04  & 0.08 $\pm$ 0.02 & 2.1 $\pm$ 1.1 & 90\\
Dec. 1$^\dagger$ & 3.51 $\pm$ 0.5 & 0.06 $\pm$ 0.13 & 0.06 $\pm$ 0.01 & 0.97 $\pm$ 0.2 & 0.58 $\pm$ 0.09 & 0.05 $\pm$ 0.02 & 0.95 $\pm$ 1.2 & 90\\
Dec. 9$^\star$ & 2.81 $\pm$ 0.4 & 0.08 $\pm$ 0.09 & 0.20 $\pm$ 0.02 & 0.86 $\pm$ 0.05 & 0.74 $\pm$ 0.04 & 0.08 $\pm$ 0.02 & 1.0 $\pm$ 0.3 & 90\\
Dec. 10$^\star$ & 2.81 $\pm$ 0.4 & -0.08 $\pm$ 0.06 & 0.20 $\pm$ 0.02 & 0.90 $\pm$ 0.06 & 0.67 $\pm$ 0.03 & 0.10 $\pm$ 0.02 & 0.8 $\pm$ 0.2 & 67 $\pm$ 4\\
\hline
\multicolumn{9}{c}{HCN} \\
\hline
Nov. 5$^\dagger$ & 2.34 $\pm$ 1.0 & 0.22 $\pm$ 0.11 & 0.08 $\pm$ 0.01 & 0.67 $\pm$ 0.05 & 0.42 $\pm$ 0.03 & 0.04 $\pm$ 0.02 & 1.4 $\pm$ 0.4 & 60 $\pm$ 5\\
Nov. 21$^\dagger$ & 3.61 $\pm$ 0.5 & -0.10 $\pm$ 0.05 & 0.23 $\pm$ 0.01 & 0.86 $\pm$ 0.04 & 0.51 $\pm$ 0.01 & 0.04 $\pm$ 0.01 & 2.8 $\pm$ 0.5 & 55 $\pm$ 3\\
Dec. 9$^\star$ & 2.81 $\pm$ 0.4 & 0.08 $\pm$ 0.02 & 0.71 $\pm$ 0.02 & 0.84 $\pm$ 0.01 & 0.63$\pm$ 0.02 & 0.07 $\pm$ 0.02 & 1.6 $\pm$ 0.1 & 90
\enddata
\vspace{4pt}
\noindent \textbf{Notes.} \\
$^\dagger$ NOEMA data, with CS from ON-OFF observations and HCN from combined ON-OFF and interferometric data.\\
$^\star$ Data from our APEX observations.\\
\sups{a} \added{H\subs{2}O production rates from SOHO/SWAN observations. Uncertainties include the formal 1$\sigma$ fitting errors reported by \citet{Combi2023}, with an additional $10\%$ fractional uncertainty added in quadrature to account for temporal mismatch and short-timescale variability.}\\
\sups{b} Mean Doppler shift (velocity first moment) for each line from NOEMA ON-OFF and APEX Spectra. \\
\sups{c} Spectrally integrated flux extracted from NOEMA ON-OFF and APEX Spectra. \\
\sups{d} Expansion velocity in the sunward hemisphere. \\
\sups{e} Expansion velocity in the anti-sunward hemisphere.\\
\sups{f} Mixing ratio with respect to H$_2$O. Reported uncertainties include an additional 10\% absolute flux-scale calibration error.\\
\sups{g} Ratio between molecular production rates for sunward and anti-sunward hemispheres.\\
\sups{h} Half-opening angle of the jet, oriented at a phase angle ($\phi$) relative to the observer and a position angle ($\psi$) in the sky plane. Values of 90$\degree$ without uncertainties correspond to a two-region model with symmetric sunward and anti-sunward hemispheres. Values with reported uncertainties are fit for the jet's half-opening angle.
\end{deluxetable*}

\section{Discussion}\label{sec:discussion}

\subsection{Velocity Differences Between CS and HCN, and Compactness of HCN}\label{subsec: spatial}
Differences in the measured gas expansion velocities of CS and HCN offer insight into the spatial origins of these species in the coma. We find that CS consistently showed higher expansion velocities compared to HCN (see Table~\ref{tab:composition_table}), particularly in the anti-sunward hemisphere. This discrepancy reflects differences in both the production of each species and how our observations sampled the coma. Gas is accelerated because of the quasi-adiabatic expansion of the flow, and within the collisional region ($R_c$ $\simeq$ 1,000 km) all species share the same expansion velocity \citep{meech2024comets}. Our interferometric visibility modeling indicates that HCN was compact and released directly from the nucleus, so its measured velocities primarily reflect this common expansion at the scales probed by NOEMA. Specifically, we obtain 2.6$\sigma$ (99$\%$ confidence) upper limits on the HCN parent scale length: $<225$ km on November 5 and $<295$ km on November 21.

CS, in contrast, was detected only in ON-OFF/single dish spectra, and its higher expansion speed can be understood in terms of its production mechanisms. As described previously, CS production in other recently studied comets is inconsistent with CS$_2$ photolysis in the inner (collisional) coma and best explained by distributed source production with $L_p\sim1,000$ km. If CS in A1 were formed by CS$_2$ photolysis, fragment excess velocities from its production would be effectively thermalized and not traceable in the outer coma at the distances sampled by the NOEMA ON-OFF/APEX beams. In contrast, the NOEMA/APEX beam sizes would be sensitive to excess velocity imparted to CS by its formation from a distributed source outside of the collisional zone, as these excesses would escape thermalization and be preserved. Thus, the higher CS line widths (and in turn, measured expansion velocities) compared to HCN in A1 are consistent with CS production from a distributed source. \added{We note that on December 9 the APEX-derived expansion velocities of CS and HCN are more similar than those reported earlier in the observing campaign. The velocity contrast is clearest in early November, when HCN was observed with NOEMA's interferometric beam and constrained to be compact. By early December, observations of HCN were obtained with the larger single-dish APEX beam, which averaged over a greater fraction of the coma and reduced sensitivity to spatial differences between nuclear and distributed sources. Additionally, increased activity near perihelion likely expanded the collisional region---further diminishing observable kinematic separation. The convergence of velocities on December 9, therefore, likely reflects differences in beam sampling and evolving coma conditions rather than contradicting a distributed source interpretation for CS.}

\subsection{Enrichment, Depletion, and Variability of Volatiles}\label{subsec: ourvar}
To place comet A1 in context, we assess whether its volatiles were enriched, depleted, or variable relative to typical Oort cloud values. These comparisons reveal strong depletion in some species, time-dependent trends in others, and help constrain the processes shaping A1's activity during its inbound leg. Our stringent $3\sigma$ upper limit on the CH$_3$OH mixing ratio in A1 ($<1.2\%$; relative to H$_2$O) is depleted compared to average values measured in Oort cloud comets \citep[$\sim\!2.2\%$;][]{DelloRusso2016}. For H$_2$CO, the 3$\sigma$ upper limit of $<0.2\%$ on its abundance is below the Oort cloud average of 0.33$\%$. In contrast, our 3$\sigma$ upper limit for the CO abundance ($<7\%$) is consistent within uncertainty with the Oort cloud average of $6.1\pm1.6\%$, ruling out A1 as a CO-enriched comet. It is important to note, however, that abundances of CO and other volatiles show substantial comet-to-comet diversity, with some comets exhibiting extreme enrichment and/or depletion \citep{Bockelee-Morvan2017}.

\added{Considering only our NOEMA/APEX measurements, the detected species CS and HCN show evidence for a net increase in their abundances toward smaller heliocentric distances, although departures from strict monotonicity are present.} CS abundances rose from \added{$0.02\pm0.01\%$ (Nov. 5; $r_H\sim1.3$ au) to $0.08\pm0.02\%$ (Nov. 21; $r_H\sim1.08$ au), then dipped to $0.05\pm0.02\%$ (Dec. 1; $r_H\sim0.93$ au) before rising sharply to $\sim\!0.09\%$ (Dec. 9-10; $r_H\sim0.8$ au). HCN abundances showed a smaller dynamic range in our data, from $\sim\!0.04\%$ (Nov. 5-21; $r_H\sim1.3$-1.08 au) to $0.07\pm0.02\%$ (Dec. 9; $r_H\sim0.8$ au).} For context, canonical averages from radio surveys are $\sim\!0.10\%$ for both CS and HCN abundances \citep{Bockelee-Morvan2017}, although we note infrared measurements typically report a higher HCN average of $\sim\!0.22\%$ \citep{DelloRusso2016}. Relative to these benchmarks, comet A1 was time-averaged depleted in both species. Nevertheless, the late-epoch increase we observed suggests that single-epoch snapshots taken closer to the Sun could partially mask longer-term signatures of cometary composition. We discuss the robustness and implications of these trends in the following section, where we place our results alongside other published measurements of comet A1.

\subsection{Comparison to Other Compositional Measurements of D/2021 A1}\label{subsec: comparison}
We first consider the iSHELL/IRTF measurements of comet A1 presented by \citet{Faggi2023}. Their observations, conducted on UT 2021 December 20 and 2022 January 8 and 9, revealed the presence of several molecules, including H\subs{2}O, HCN, C\subs{2}H\subs{2}, NH\subs{3}, NH\subs{2}, C\subs{2}H\subs{6}, CH\subs{4}, H\subs{2}CO, CO, OCS, and HCl. Their mixing ratios were normalized to iSHELL $Q(\mathrm{H_2O})$ at those dates, whereas ours used SOHO/SWAN $Q(\mathrm{H_2O})$ from \citet{Combi2023} at earlier epochs, so differences in reported abundances may have arisen both from the choice of reference water production rate and from the fact that the observations were not simultaneous. They retrieved HCN mixing ratios of $0.10 \pm 0.01\%$ on December 20 and $0.11 \pm 0.01\%$ on January 9. These values are elevated compared to our weighted average HCN mixing ratio of \added{$0.05 \pm 0.01\%$}. However, since they measured in the infrared (IR), their HCN values may have reflected the known trend of IR-derived HCN abundances being approximately twice as high as those from radio measurements \citep{Bockelee-Morvan2017}.

For H\subs{2}CO, \cite{Faggi2023} reported a mixing ratio of $0.14 \pm 0.01\%$ on January 8, which is compatible with our upper limit of $<0.2\%$. Similarly, their CO mixing ratio of $1.03 \pm 0.01\%$ on January 8 aligns well with our upper limit of $<7\%$. For CH\subs{3}OH, they established upper limits of $<0.07\%$ on January 8 and $<0.22\%$ on January 9, both of which were encompassed within our reported upper limit of $<1.2\%$. The stronger depletion of CH\subs{3}OH, relative to other volatile species like CO, CH\subs{4}, and C\subs{2}H\subs{6} was interpreted as evidence of interstellar/solar nebular chemistry signatures in A1's ices \citep{Faggi2023}. Moreover, the detections of HCl and strong detections of OCS further supported the idea of interstellar origin for A1's ices, as both molecules are considered to be preferentially formed via solid-phase processes in interstellar chemistry \citep{Faggi2023}.

In another study of comet A1, \citet{Biver2024} used the IRAM 30-m radio telescope to observe the comet during November-December 2021, spanning heliocentric distances from 1.22 to 0.76 au---similar to our coverage. In contrast to our non-detections across the campaign, they detected CH$_3$OH and H$_2$CO in November and December. Our most direct comparison is on December 9, when we both observed comet A1 at similar heliocentric distances. \added{From our APEX spectrum, we derived a $3\sigma$ upper limit for CH$_3$OH production of $<\!3.9\times10^{26}~\mathrm{molecules~s^{-1}}$ and $<\!6.4\times10^{25}~\mathrm{molecules~s^{-1}}$ for H$_2$CO using an H$_2$O production rate of $(2.81 \pm 0.4)\times10^{28}~\mathrm{molecules~s^{-1}}$. \citet{Biver2024} adopted a somewhat higher H$_2$O production rate of $4\times10^{28}\mathrm{~molecules~s^{-1}}$ for December 8-13 and reported a CH$_3$OH production rate of $(3.7\pm0.7)\times10^{26}~\mathrm{molecules~s^{-1}}$ on December 9 and H$_2$CO production rate of $(9.5\pm2.8)\times10^{25}~\mathrm{molecules~s^{-1}}$ on December 8. Because molecular mixing ratios are defined relative to the assumed H$_2$O production rate, differences in $Q(\mathrm{H_2O})$ between studies can shift the absolute normalization of reported production rates and abundance ratios. For CH$_3$OH, our 3$\sigma$ upper limit of $<\!3.9\times10^{26}~\mathrm{molecules~s^{-1}}$ on December 9 encompasses the value reported by \citet{Biver2024}, $(3.7\pm0.7)\times10^{26}~\mathrm{molecules~s^{-1}}$, and therefore is not in tension. However, for H$_2$CO, the production rate reported by \citet{Biver2024} is slightly higher than our formal upper limit. If, however, we examine the mixing ratio relative to the adopted water production rates, our constraint corresponds to $<\!0.2\%$ relative to H$_2$O, while \citet{Biver2024}'s December 8 value corresponds to $0.24 \pm 0.07\%$. The lower bound of their reported range is consistent with our 3$\sigma$ limit, indicating no statistically significant discrepancy.}

\begin{figure*}
    \centering
    \includegraphics[width=1\textwidth]{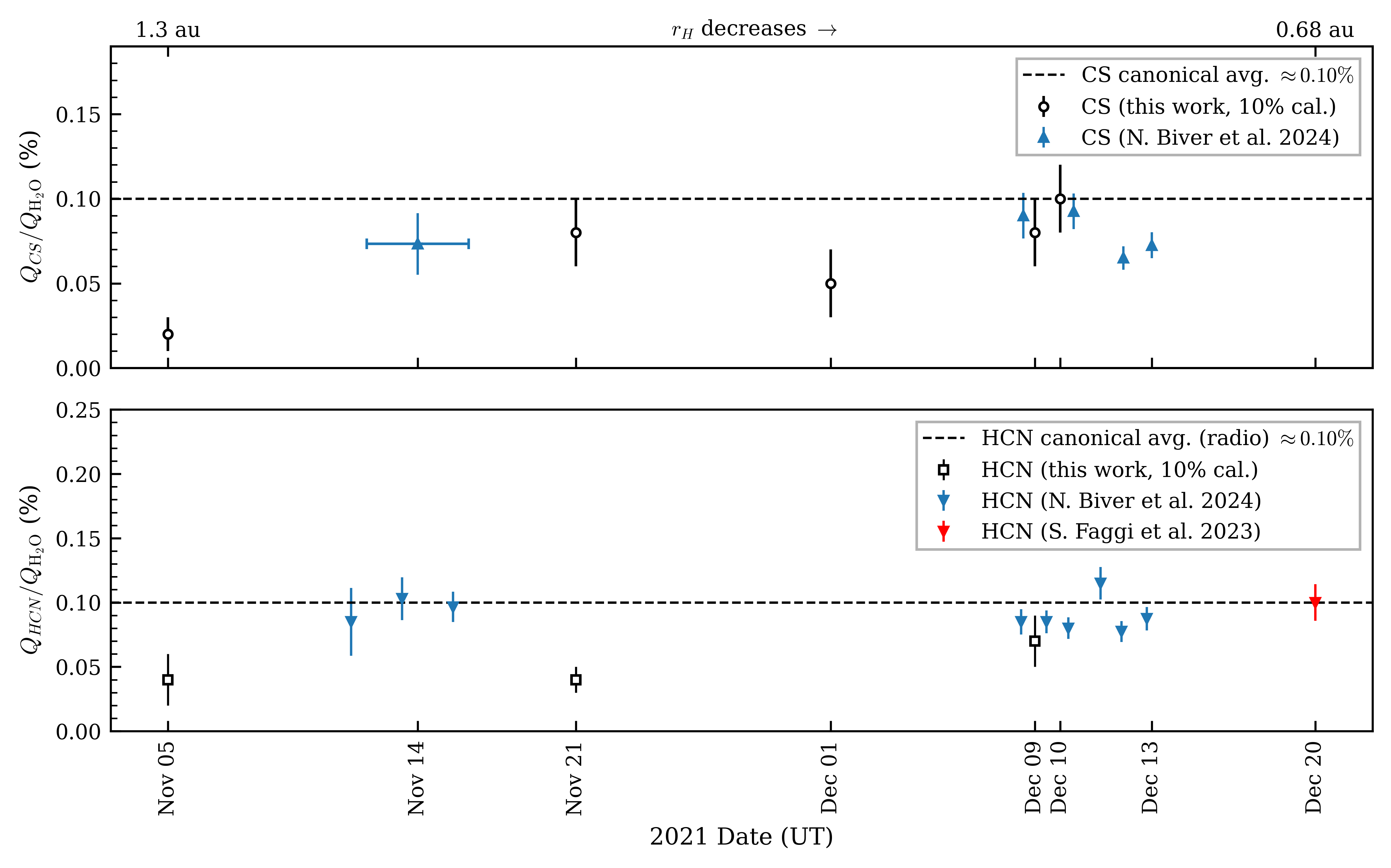}
    \caption{Pre-perihelion time evolution (decreasing $r_H$) of the CS (top) and HCN (bottom) mixing ratios relative to H$_2$O in comet A1. Black open symbols show results from this work. Blue colored, solid symbols show contemporaneous measurements adopted from \citet{Biver2024} while the red colored, solid symbol shows measurements from \citet{Faggi2023}. Shown error bars represent 1$\sigma$ uncertainties that include an additional 10$\%$ contribution from the absolute flux-scale calibration to account for comparison between differing facilities. Dashed lines represent radio canonical average abundances from \citet{Bockelee-Morvan2017}.}
    \label{fig:CS_HCN_evolution}
\end{figure*}

For other species, \citet{Biver2024} reported mixing ratios for HCN of $0.09 \pm 0.01\%$, modestly higher than our campaign's weighted average of \added{$0.05 \pm 0.01\%$} over a comparable $r_H$ range, and CS mixing ratios of $0.06 \pm 0.01\%$, statistically consistent with our weighted average of \added{$0.05 \pm 0.01\%$} and falling within the spread of our measurements (Table~\ref{tab:composition_table}). Although daily mixing ratios were not tabulated in their paper, we combined their reported daily/weekly molecular production rates with their quoted H$_2$O production to reconstruct an approximate temporal evolution of volatile mixing ratios. Figure~\ref{fig:CS_HCN_evolution} shows this comparison, with our retrieved mixing ratios plotted alongside the contemporaneous adopted values from \citet{Biver2024} and pre-perihelion HCN values from \citet{Faggi2023} (CS in the top panel, HCN in the bottom). 

\begin{figure*}
    \centering
    \includegraphics[width=1\textwidth]{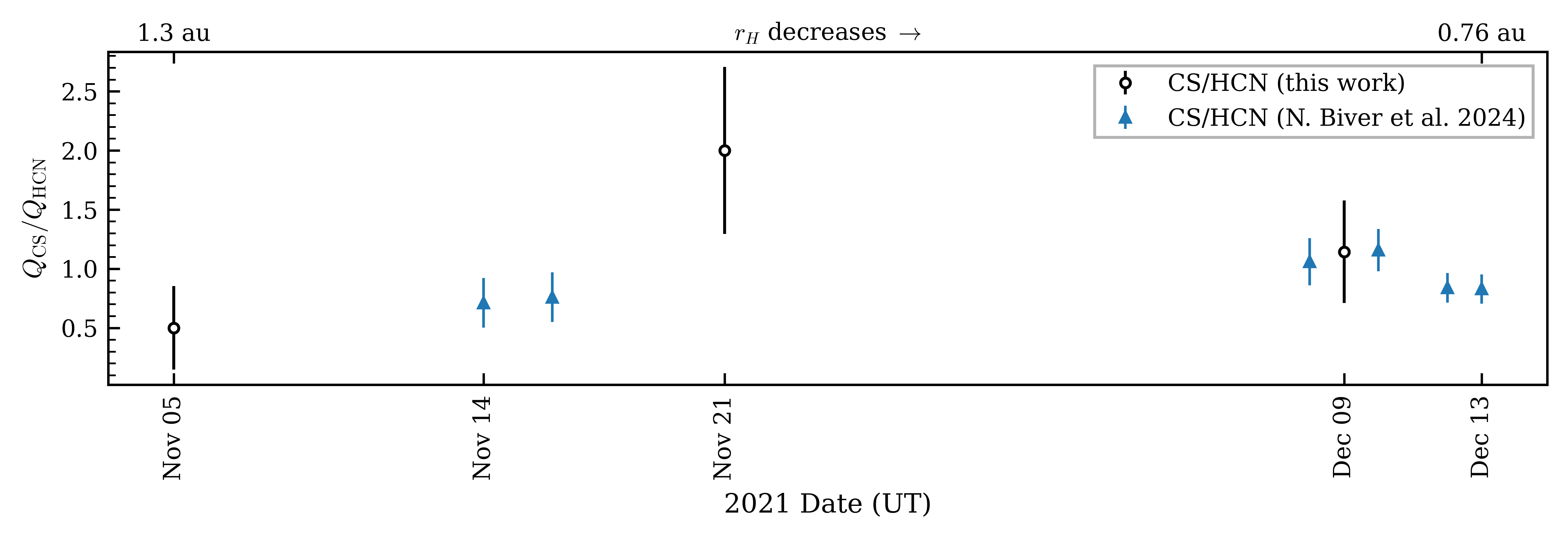}
    \caption{Pre-perihelion time evolution (decreasing $r_H$) of the CS/HCN mixing ratio in comet A1. Because the ratio does not rely on normalization to H$_2$O, it provides a water-independent comparison of relative abundance variability. Black open symbols show results from this work. Blue colored, solid symbols show contemporaneous measurements adopted from \citet{Biver2024}. In November, for \citet{Biver2024}, the CS abundance is taken as the weekly-average value reported for Nov 13–16, and this same CS value is paired with the two HCN measurements within that window (Nov 14 and Nov 16) to form CS/HCN. Shown error bars represent 1$\sigma$ uncertainties that include an additional 10$\%$ contribution from the absolute flux-scale calibration to account for comparison between differing facilities.}
    \label{fig:CS_over_HCN_evolution}
\end{figure*}

For CS, incorporating the \citet{Biver2024} dataset supports the trend discussed in Section~\ref{subsec: ourvar}: \added{a suggestive increase in abundance with decreasing $r_H$, from $\sim\!0.02\%$ near 1.3 au in early November to $\sim\!0.08$-$0.10\%$ by 0.8 au in December. Our early-November observation anchors the lower baseline, and a net increase toward smaller $r_H$ remains evident once the denser coverage is included, although the significance of a strictly monotonic slope is modest once cross-facility uncertainties are included}. For HCN, a broad monotonic increase in abundance with decreasing $r_H$ is not as apparent once the \citet{Biver2024} values are included. Our measurements alone suggest a gradual rise from $\sim\!0.04\%$ in early November to $\sim\!0.07\%$ by December 9, but the combined record occupies a higher baseline of $\sim\!0.08\%$ with modest variability in mid-December that does not track $r_H$. \added{We note that our HCN abundances are systematically lower than those reported by \citet{Biver2024}. In their analysis of A1, allowing for a distributed source with a parent scale length of order 350~km increases the inferred $Q(\mathrm{HCN})$ by $15-35\%$ relative to a purely nuclear source ($L_p  = 0$~km). In contrast, our visibility-based modeling disfavors extended production, placing 99$\%$ confidence upper limits of $L_p < 255$~km (Nov. 5) and $L_p < 295$~km (Nov. 21). The baseline offset between the two datasets therefore likely stems from differences in source distribution assumptions rather than intrinsic compositional discrepancies. Like HCN's variability in mid-December, we note similar variability is seen in CS around the same time; however, in line with \citet{Biver2024}'s finding of no short-term variability in A1's absolute outgassing, we interpret this variability as dominated by calibration and $Q(\mathrm{H_2O})$ effects.} \citet{Biver2024} computed molecular production using block estimates of $Q(\mathrm{H_2O})$ ($2, 3,$ and $4\times10^{28}$ s$^{-1}$ for Nov. 12, Nov. 13-16, and Dec. 8-13, respectively), so unseen day-to-day changes in $Q(\mathrm{H_2O})$ can imprint apparent structure in $Q(\mathrm{X})/Q(\mathrm{H_2O})$ without requiring changes in $Q(\mathrm{X})$. The $0.10 \pm 0.01\%$ value on December 20 reported by \citet{Faggi2023} is compatible with this baseline picture.

\added{To test whether this apparent heliocentric behavior reflects differential evolution between CS and HCN, we examined the CS/HCN ratio directly (Figure~\ref{fig:CS_over_HCN_evolution}). Because this ratio does not depend on normalization to $Q(\mathrm{H_2O})$, it is insensitive to non-simultaneous water measurements and differing assumptions between studies. Within uncertainties, CS/HCN is consistent with a constant value over the pre-perihelion interval. This indicates that any heliocentric increase inferred for CS is modest and not clearly distinguishable from the evolution seen in HCN, rather than reflecting strong differential compositional evolution between the two species.}

Taken together, A1's inbound record suggests that heliocentric trends highlighted in prior studies (e.g., \citealt{Irvine2004, Fray2006, DiSanti2016, DelloRusso2016, Cordiner2017}) were beginning to emerge in a nucleus already destabilizing. \added{The modest increase of CS abundance toward smaller $r_H$ is consistent with production from a distributed, refractory source (\citealt{Cottin2008} and references therein), whereas HCN shows at most a modest baseline shift and no statistically robust monotonic dependence on $r_H$, consistent with predominantly near-nucleus release.} This agrees with our spatial interpretation in Section~\ref{subsec: spatial}. Noticeable short-term variability in mixing ratios is more apparent for both species later in December, coincident with recurrent outbursts and the onset of fragmentation \citep{Crovisier2021, Jewitt_2023}. \added{Because ratios reference H$_2$O, the variability likely reflects changes in the adopted water production rate, with any intrinsic shifts during disintegration superposed at a level that is difficult to isolate (e.g., transient exposure of fresher material).} Therefore, we interpret the mid-December variability as departures coincident with reported disruption superposed on the smoother heliocentric evolution seen earlier. Overall, A1 bridges thermally driven increases in mixing ratios at small $r_H$ with the volatile-rich, disruption-linked activity that can dominate as long-period comets approach disintegration.

\section{Conclusion}
Through this coordinated pre-perihelion monitoring campaign with NOEMA and APEX, we have established a baseline characterization of volatile release in comet A1 between 1.3 and 0.80 au. Coupled with our \texttt{SUBLIME} radiative transfer modeling, we placed stringent (3$\sigma$) upper limits on CH$_3$OH, H$_2$CO, and CO abundance, and we retrieved expansion velocities and mixing ratios for the key species HCN and CS. The retrieved gas expansion velocities showed a systematic difference between species: CS generally exhibited larger velocities compared to HCN. This velocity difference is consistent with their spatial origins. Our interferometric visibility modeling indicates that HCN production was compact and originated from nucleus sources, with parent scale lengths constrained to $<$300 km. In contrast, CS is best explained as a distributed source being produced over more extended distances, where fragment excess velocities were preserved and ON-OFF/single-dish beam sampling emphasized the outer coma. Together, the preservation of fragment excess velocities in extended CS production and the contrasting spatial sensitivity of interferometric versus ON-OFF/single-dish observations provide a consistent explanation for the systematically larger expansion velocities retrieved for CS compared to HCN. 

\added{Temporally, we find that CS exhibits a modest increase in mixing ratio toward smaller heliocentric distances, consistent with distributed release becoming more efficient under increasing solar heating. HCN, in contrast, shows comparatively limited evolution over the pre-perihelion interval, remaining within the $0.05-0.08\%$ range spanned by the combined datasets and lacking a statistically robust monotonic dependence on $r_H$}. Noticeable variability in both species coincided with the mid-December outbursts and the onset of nucleus fragmentation \citep{Crovisier2021, Jewitt_2023}. These departures are consistent with disruption intermittently exposing fresh material and/or altering $Q(\mathrm{H_2O})$ (thereby affecting mixing ratios), but they do not by themselves demonstrate a sustained compositional shift. 

These patterns in CS and HCN mixing ratios indicate that A1's inbound volatile evolution may not be explained by increasing solar insolation alone. \added{CS exhibits a month-scale baseline increase consistent with a thermal response, whereas day-to-day departures that appear in mid-December coincide with reported disruption,} suggesting an additional role for nucleus instability. Given the current cadence and inter-facility differences in beam size, calibration, and adopted $Q(\mathrm{H_2O})$, cleanly separating long-term trends from short-term activity is challenging. The most conservative interpretation is that both increased solar insolation and fragmentation contributed to A1's volatile behavior, with insolation setting the baseline evolution and disruption superposing brief departures late in the apparition. Broader temporal coverage and higher-cadence, multi-facility monitoring in future campaigns will be needed to more firmly disentangle these effects.

\section{Data Availability}\label{sec:data_avail}
\added{Data from the IRAM NOEMA interferometer under project code S21AA are publicly accessible through the IRAM Data Archive at \url{https://iram-institute.org/science-portal/data-archive/}. Data from the APEX telescope under project number 0108.F-9322 are publicly accessible through the APEX science archive facility at \url{https://archive.eso.org/wdb/wdb/eso/apex/form}.}

\begin{acknowledgments}
This work was based on observations carried out under project number S21AA with the IRAM NOEMA Interferometer, supported by INSU/CNRS (France), MPG (Germany), and IGN (Spain). This work also used data acquired with the Atacama Pathfinder Experiment (APEX) under project number 0108.F-9322. APEX is a collaboration between the Max-Planck-Institut fur Radioastronomie, the European Southern Observatory, and the Onsala Space Observatory. 

This work was supported by the NASA Solar System Observations program (grants 80NSSC24K1324: N.X.R., S.N.M., M.A.C.; 80NSSC22K1401: B.P.B., N.D.R.), the Planetary Science Division Internal Scientist Funding Program through the Fundamental Laboratory Research (FLaRe) work package (T.N.P., N.X.R., S.N.M., M.A.C., S.B.C.), as well as the NASA Astrobiology Institute through the Goddard Center for Astrobiology (proposal 13-13NAI7-0032; S.N.M., M.A.C., S.B.C.). T.N.P. additionally acknowledges support from the John C. Mather Nobel Scholarship.   
\end{acknowledgments}

\software{
GILDAS/CLASS (\citealt{Pety2005, team2013gildas}; \url{https://www.iram.fr/IRAMFR/GILDAS/}),
lmfit \citep{newville2016lmfit},
vis-sample \citep{Loomis2018},
SUBLIMED \citep{cordiner2023gas, Cordiner_2022}.
}

\newpage
\bibliography{LEONARD}{}
\bibliographystyle{aasjournalv7}



\end{document}